\documentclass[11pt,a4paper]{article}

\usepackage[dvipsnames]{xcolor}
\usepackage[margin=0.8in]{geometry}
\usepackage{amsmath,amssymb,mathtools,bm}
\usepackage{physics}
\usepackage{microtype}
\usepackage{enumitem}
\usepackage{hyperref}
\usepackage[numbers,sort&compress]{natbib}
\hypersetup{colorlinks=true, linkcolor=blue!50!black, urlcolor=blue!50!black}
\usepackage{authblk} 

\usepackage{url} 

\numberwithin{equation}{section}
\setlist[itemize]{noitemsep, topsep=2pt}

\makeatletter
\newcommand{\email}[1]{\gdef\@email{#1}}
\newcommand{\affiliation}[1]{\gdef\@affiliation{#1}}
\let\@email\@empty
\let\@affiliation\@empty

\renewcommand{\maketitle}{%
  \begin{center}
    {\LARGE\bfseries \@title \par}%
    \vskip 1.5em
    {\large \@author \par}%
    \vskip 0.5em
    {\normalsize \@affiliation \par}%
    \vskip 0.25em
    {\normalsize \texttt{\@email} \par}%
    \vskip 1em
    {\normalsize \@date \par}%
  \end{center}%
  \vskip 2em
}
\makeatother

\begin{document}

\title{A pathway to non-perturbative Quantum Affine Gravity}
\author{Agustín Silva}
\email{agustin.silva@ru.nl}
\affiliation{High Energy Physics Department, Institute for Mathematics, Astrophysics, and Particle Physics, Radboud University, Nijmegen, The Netherlands.}
\date{\today}

\maketitle
\begin{abstract}
   We explore a new route toward a non-perturbative quantization of gravity based on a purely affine formulation, where the affine connection is the fundamental field and the metric, when it exists, emerges as a derived quantity. Starting from the Palatini formulation of General Relativity, we recall how an equivalent Eddington-type purely affine action arises at the classical level under mild assumptions. A key feature for the non-perturbative program is that, in the pure gravity case, for a positive cosmological constant, the action is bounded below, allowing one to define a well-posed statistical ensemble of connections. We discretize this theory on a fixed hypercubic lattice and construct the corresponding partition function, including torsionful and torsionless ensembles. We provide an open C++ Monte Carlo implementation that can simulate these ensembles in arbitrary dimension, and we present proof-of-principle results in two dimensions.
\end{abstract}

\tableofcontents
\section{Introduction}
\label{sec:intro}

Despite many decades of effort, a satisfactory quantum theory of gravity remains elusive.  
The perturbative quantization of the Einstein--Hilbert action is non-renormalizable: already at one loop the presence of matter spoils renormalizability, and at two loops even pure gravity develops non-renormalizable ultraviolet divergences \cite{tHooft:1974toh,Goroff:1985sz}.
This has motivated a variety of non-perturbative approaches, both in the continuum and on the lattice.

While a common starting point is to assume that the metric is the fundamental degree of freedom of gravity, in this work we explore a different starting point: a purely \emph{affine} description of gravity, in which the dynamical variable is the affine connection and the metric, when it exists at all, is an emergent, derived object.
The main novelty of this paper is to bring this purely affine (Eddington-type) formulation into a concrete non-perturbative setting by constructing an explicit lattice statistical model for the connection, together with a working Monte Carlo implementation.

The purely affine viewpoint goes back to Eddington's proposal of an action built from the determinant of the (symmetric) Ricci tensor \cite{Eddington:1924}, Einstein's theory of the affine field \cite{Einstein:1923}, and Schr\"odinger's extensive study of affine theories \cite{Schrodinger:1950}.  
In the modern literature, purely affine and metric-affine formulations have been analyzed in detail, both at the classical and quantum level, giving birth to Eddington-type and Ricci-based theories such as Eddington-inspired Born-Infeld gravity (EiBI), and other extensions to General Relativity (GR) \cite{astronomy3010004,Kijowski:1978,Kijowski:2007,Ferraris:1981,Poplawski:2007,Poplawski:2008,CervantesCota:2016,Martellini:1984gy,KijowskiTulczyjew:1979,Vollick:2004,Banados:2010ix,Avelino:2012ge,BeltranJimenez:2017,Chen:2016,Olmo:2011uz,Hehl:1994ue,Sotiriou:2006qn,Kalmykov:1994fm,Kalmykov:1994yj,Martellini:1983tx,Martellini:1984ec}.  
These models show that determinant-type actions can be classically equivalent to GR in suitable regimes and parameter ranges, as we will later re-derive for clarity.

Beyond this classical equivalence, there is also an aesthetic motivation for taking the connection as fundamental.  
All Standard Model interactions are mediated by excitations of gauge connections (photons, $W^{\pm}$, $Z$, gluons).  
From this perspective, it is natural to ask whether gravity might also be described in terms of excitations of a connection rather than of a metric field.  
In a purely affine formulation, a ``graviton'' would be replaced, at least heuristically, by an excitation of the affine connection itself, a ``connecton''. 
The framework explored here is a step towards making such a picture precise in a non-perturbative setting.

Our starting point is the Palatini formulation of GR in $D$ spacetime dimensions \cite{Ferraris:1982wci}, where we re-derive a purely affine theory that is equivalent to GR. 
For $D>2$, and under mild assumptions such as having an arbitrarily small, non-vanishing positive cosmological constant, the metric can be eliminated algebraically in terms of the symmetric part of the Ricci tensor (and the matter sources if present).  
This yields a purely affine action of Eddington-type that is classically equivalent to GR. Thus, via this algebraic mechanism, the affine and metric descriptions carry the same local propagating degrees of freedom.
A crucial property for our purposes is that, once the algebraic solution for the metric is inserted back into the action, the resulting Lagrangian density has a \emph{definite sign}, determined by the sign of the cosmological constant.  
For a positive cosmological constant, the action is bounded below and strictly positive.  
This contrasts sharply with the usual metric formulation given by the Einstein--Hilbert action, whose unboundedness due to the conformal mode is a long-standing obstacle to defining a gravitational path integral.  
\emph{The bounded nature of the purely affine formulation is the core motivation for this work}, because it allows one to define a non-perturbative statistical weight directly from the action.

In this paper we construct a non-perturbative, lattice formulation of the purely affine Eddington-type action and study it as a statistical field theory of the affine connection. 
We discretize spacetime on a fixed hypercubic lattice, approximate the relevant quantities by finite differences (and, as an alternative, via gravitational Wilson loops \cite{Wilson:1974sk}), and define a statistical partition function with Boltzmann weight $\exp(-S)$, where $S$ is the lattice version of the derived purely affine action. 
Because the action has a definite sign, this defines a well-posed Gibbs measure over affine connections, provided we specify the ensemble and handle redundant degrees of freedom, as we will do later on. 
The lattice explicitly breaks diffeomorphism invariance, alleviating some technical problems, but creating others, such as the need to study the recovery of diffeomorphism invariance in a suitable lattice-spacing-to-zero limit.

Our main results can be summarized as follows:
\begin{enumerate}
  \item On a hypercubic lattice, we formulate torsionful and torsionless ensembles of affine connections, and construct the corresponding statistical partition functions.
  \item We implement a local Metropolis algorithm for the affine connection and provide a publicly available C++ code \cite{silva_lqagmc_2025} capable of simulating these ensembles in arbitrary dimension $D$.
        As a first numerical test, we perform Monte Carlo simulations in $D=2$ of the torsionless ensemble on lattices of three different sizes, defining and computing the expectation values of two different diffeomorphism-invariant quantities that characterize different volume measures of the emergent geometries.  
        We find that the resulting statistical system is well-behaved: the distribution of the measured quantities depends smoothly on the parameters of the theory, with no signs of instabilities or pathologies in the explored regime.
\end{enumerate}

The scope of this work is to take the first steps into this new research avenue and to establish a concrete formulation and computational baseline from which more refined ensembles and detailed studies can be developed.

We provide specific ensembles to define the partition function, as well as a code to simulate the systems in $D$ dimensions, but we do not dive into constraining such ensembles so that specific metric properties emerge from them, as we will discuss later on. 
Since these physical constraints are not yet imposed (and their implementation would most likely require substantial additional work), in this manuscript, we restrict the numerical analysis to the simplest possible scenario. 
As well, we do \emph{not} claim to have constructed a Lorentzian, unitary quantum theory of gravity.  
We work entirely within a statistical framework and do not address the existence of a Wick rotation.

Treating gravity as a statistical field theory is standard practice in the early stages of development of an approach, and has not prevented the community from exploring such models even when the precise relation to a Lorentzian, unitary theory is not fully under control. 
One prominent example is the Asymptotic Safety scenario, originally suggested by Weinberg \cite{Weinberg:1979}, in which gravity is described by a quantum field theory whose renormalization group flow approaches a non-trivial ultraviolet fixed point. 
While very recently there has been progress towards Lorentzian calculations \cite{DAngelo:2023tis,DAngelo:2023wje,DAngelo:2025yoy,Pawlowski:2023gym,Saueressig:2025ypi,Fehre:2021eob}, in the last three decades most progress has been achieved using Functional Renormalization Group (FRG) computations that do not assume a particular Wick rotation of the effective field theories studied, nor a fixed signature for the quantized metrics. While some computations do fix the signature of the full metric, most split the metric into background and fluctuation and only fix the signature of the background, but not of the full metric.
In this sense, while substantial evidence for such a fixed point has been accumulated in FRG studies of ``quantum Einstein gravity'' and extensions \cite{Reuter:1996cp,Reuter:2001ag,Percacci:2007,Reuter:2012,Eichhorn:2018yfc,Dona:2013qba,Bonanno:2015fga,Falls:2018ylp,Bonanno:2020bil,Knorr:2022dsx,Niedermaier:2006wt,Eichhorn:2022gku,Morris:2022btf,Martini:2022sll,Pawlowski:2023gym,Platania:2023srt,Reichert:2020mja,Basile:2024oms}, the setting where many of those results are derived can be qualified as purely statistical.

On the lattice, a similar path has been pursued. 
Early work constructing gravitational ensembles includes Regge calculus and, from it, studies of simplicial geometries within Euclidean Dynamical Triangulations (EDT) \cite{Ambjorn:1992EDT,Agishtein:1992EDT,Coumbe:2014nea,Ambjorn:2013eha,Schiffer:2025cqc}. 
While EDT is a statistical framework with no clear notion of a time direction and no explicit implementation of a Wick rotation, its exploration led to more refined ensembles with a notion of proper-time foliation and a concrete Wick rotation, within Causal Dynamical Triangulations (CDT) \cite{Ambjorn:2000dv,Ambjorn:2001cv,Ambjorn:2012jv,Loll:2019rdj,Ambjorn:2024CDT,Ambjorn:2020rcn,Ambjorn:2024as,Ambjorn:2007jv,Ambjorn:2005db,Ambjorn:2005qt,Ambjorn:2024bud,Ambjorn:2022dvx}. 
As in any lattice formulation of a QFT, a suitable quantum scale-invariant regime in parameter space is usually needed to find a continuum limit (sometimes in the same form as in Asymptotic Safety, with a UV fixed point). 
This is the case in CDT, where so far promising results have been found \cite{Ambjorn:2014gsa,Ambjorn:2024as}.

A scale-invariant limit in parameter space would also be required to define a continuum limit of the statistical model studied in this paper. 
While some contact between affine gravity and FRG studies exists to date such as \cite{Knorr:2020bjm,Percacci:2020bzf,Pagani:2015ema,Gies:2022ikv}, this has not been analyzed to the extent that metric formulations have been, and to the best of my knowledge, a lattice counterpart such as the one studied in this paper has never been explored. 

In that sense, our goal here is modest: to show that the purely affine Eddington-type action admits a concrete, statistical, non-perturbative lattice realization, with well-behaved Monte Carlo dynamics, and to provide a computational framework from which more refined ensembles and detailed studies can be developed. 
This work should not be viewed as a complete realization of the vast amount of work that this new line of research opens, but as a solid starting point.

The rest of the paper is organized as follows.  
In Sec.~\ref{sec:palatini} we review the Palatini formulation, adding simple matter content for generality, and re-derive the purely affine Eddington-type action that is classically equivalent to GR, including a discussion of the conditions for the existence of an emergent metric and the special role of two dimensions.  
In Sec.~\ref{sec:lattice-ricci} we discuss the lattice approximation for the partition function of this affine theory, and define the torsionful and torsionless statistical ensembles, including a discussion of gauge fixing and path integral measure definitions.  
In Sec.~\ref{sec:numerics} we test the developed code in the simplest toy model accessible within Monte Carlo simulations, that of a two-dimensional manifold, and provide concrete diffeomorphism-invariant quantities that can be computed in the simulations, as well as specific numerical results.  
We conclude in Sec.~\ref{sec:concl} with a summary and an outlook on extensions to higher dimensions, improved discretizations, and the question of continuum and Lorentzian limits.

\section{Affine formulation of General Relativity}
\label{sec:palatini}

As stated before, we will explore the first steps for quantizing gravity under the assumption that the metric is not fundamental, but instead the connection is. It has long been known that GR can be written in a purely affine form, as Eddington already observed almost a century ago \cite{Eddington:1924}. In this section, we briefly review this construction in a form adapted to our purposes, emphasizing the conditions under which it is equivalent to the usual metric formulation.

\subsection{Palatini formulation and two algebraic routes}

We start from the Einstein--Hilbert action in Palatini form, and for generality of the discussion, we include a minimally coupled real scalar field with self-interaction potential $V(\phi)$, in $D$ spacetime dimensions,
\begin{equation}
  S[g,\Gamma,\phi] \;=\; \frac{1}{16\pi G}\int d^D x\,\sqrt{|g|}\;g^{\mu\nu} R_{\mu\nu}(\Gamma)
  \;+\;\int d^D x\,\sqrt{|g|}\,\Big(-\tfrac12 g^{\mu\nu}\,\partial_\mu\phi\,\partial_\nu\phi - V(\phi)\Big),
  \label{eq:palatini-action}
\end{equation}
where $R_{\mu\nu}(\Gamma)$ is the Ricci tensor of an \emph{a priori} independent affine connection $\Gamma^\lambda{}_{\mu\nu}$.

Varying with respect to $\Gamma^\lambda{}_{\mu\nu}$ and restricting for simplicity to torsionless connections, one finds
\begin{equation}
  \nabla_\lambda\!\big(\sqrt{|g|}\,g^{\mu\nu}\big)
  -\delta^\nu{}_\lambda\,\nabla_\sigma\!\big(\sqrt{|g|}\,g^{\mu\sigma}\big)=0,
\end{equation}
which implies metric compatibility,
\begin{equation}
  \nabla_\lambda g_{\mu\nu}=0,
\end{equation}
and hence that $\Gamma$ is the Levi--Civita connection of $g_{\mu\nu}$. Substituting this solution back into \eqref{eq:palatini-action} yields the usual metric formulation of GR with a scalar field.

For our purposes, it is useful to emphasize that the Palatini system admits \emph{two} algebraic routes:
\begin{itemize}
  \item the standard one, just recalled, in which one solves the connection equation $\delta_\Gamma S=0$ and eliminates $\Gamma$ in favour of $g_{\mu\nu}$;
  \item an alternative route, in which one treats the metric as an auxiliary field and solves the metric equation $\delta_g S=0$ algebraically in terms of the affine curvature and matter sources, eliminating $g_{\mu\nu}$ in favour of $R_{(\mu\nu)}(\Gamma)$.
\end{itemize}
In the second case, the affine connection is taken as the fundamental variable, and the metric, when it exists, becomes a derived quantity.

\subsection{Eliminating the metric}

Varying \eqref{eq:palatini-action} with respect to $g_{\mu\nu}$ gives
\begin{equation}
  R_{(\mu\nu)}(\Gamma)-\tfrac12 g_{\mu\nu}g^{\alpha\beta}R_{(\alpha\beta)}(\Gamma) \;=\; 8\pi G\,T_{\mu\nu},
  \qquad \, \text{with}\,\,\,\,
  T_{\mu\nu}=\partial_\mu\phi\,\partial_\nu\phi - g_{\mu\nu}\Big(\tfrac12 g^{\rho\sigma}\partial_\rho\phi\,\partial_\sigma\phi + V(\phi)\Big),
  \label{eq:eqmetricalgebraic}
\end{equation}
where parentheses denote symmetrization. In the Palatini setting $R_{(\mu\nu)}(\Gamma)$ is built purely from the affine connection, and \eqref{eq:eqmetricalgebraic} is an \emph{algebraic} equation for the metric.

A convenient way to see how $g_{\mu\nu}$ can be solved for is to move the $\partial_\mu\phi\,\partial_\nu\phi$ term to the left-hand side and collect everything proportional to $g_{\mu\nu}$ on the right-hand side. Defining
\begin{equation}
  P_{\mu\nu}\;\equiv\;R_{(\mu\nu)}(\Gamma)-8\pi G\,\partial_\mu\phi\,\partial_\nu\phi,
  \label{eq:P_def}
\end{equation}
one finds that \eqref{eq:eqmetricalgebraic} is equivalent to a proportionality relation of the form
$P_{\mu\nu}=C\,g_{\mu\nu}$, where the scalar coefficient $C$ depends only on the trace of the equation.
Taking the trace and using $D\neq 2$ to avoid degeneracy, the trace algebra fixes this coefficient uniquely to
$C=\tfrac{16\pi G}{D-2}\,V(\phi)$.
Therefore, for $D>2$ and for potentials $V(\phi)$ that do not vanish, $g_{\mu\nu}$ is proportional to $P_{\mu\nu}$ and can be reconstructed by inversion.\footnote{See e.g.~\cite{Ferraris:1981,Kijowski:1978,Poplawski:2007,CervantesCota:2016} for more detailed discussions in the purely affine and metric--affine literature.} A convenient way to write the result is
\begin{equation}
  g_{\mu\nu}
  \;=\; \frac{D-2}{16\pi G\,V(\phi)}\Big(R_{(\mu\nu)}(\Gamma)-8\pi G\,\partial_\mu\phi\,\partial_\nu\phi\Big),
  \label{eq:metric-from-ricci}
\end{equation}
which is only valid when the RHS is invertible.

This expression makes it clear that an emergent metric can be reconstructed provided the following conditions hold:
\begin{enumerate}
  \item $D\neq 2$, so that the trace of \eqref{eq:eqmetricalgebraic} does not degenerate;
  \item the scalar potential $V(\phi)$ does not vanish in the region of field space explored by the theory, so that the overall scale of $g_{\mu\nu}$ is well defined;
  \item the symmetric tensor $P_{\mu\nu}\;=\;R_{(\mu\nu)}(\Gamma)-8\pi G\,\partial_\mu\phi\,\partial_\nu\phi$ is non--degenerate, so that the resulting metric is invertible.
\end{enumerate}
If any of these conditions fails, the algebraic relation cannot be inverted and the metric description breaks down. In particular, the $D\to 2$ limit of \eqref{eq:metric-from-ricci} is singular, and no metric can be defined in that case, even though the purely affine action itself has a smooth $D\to 2$ limit, as we will see later.

\subsection{Purely affine action and sign properties}

Substituting \eqref{eq:metric-from-ricci} back into the action \eqref{eq:palatini-action}, one obtains a purely affine action depending only on $\Gamma$ and $\phi$,
\begin{equation}
  S_{\text{aff}}[\Gamma,\phi]
  \;=\; \frac{(D-1)(D-2)^{\frac{D-2}{2}}}{(16\pi G)^{\frac{D}{2}}}
  \int d^D x\;
  \sqrt{|\det\!\Big(R_{(\mu\nu)}-8\pi G\,\partial_\mu\phi\,\partial_\nu\phi\Big)|}\;
  |V(\phi)|^{-\frac{D}{2}}\,V(\phi)\, ,
  \label{eq:edd-matter}
\end{equation}
where we dropped explicit dependence on the coordinates and on $\Gamma$ for compactness.
In the pure gravity case, obtained by setting $\phi=0$ and $V(\phi)=\frac{2\Lambda}{16 \pi G}$, this reduces to Eddington's purely affine action
\begin{equation}
  S_{\text{Edd}}[\Gamma]
  \;=\; \frac{(D-1)(D-2)^{\frac{D-2}{2}}\,2\Lambda}{(16\pi G)^{\frac{D+2}{2}}}\, \Big|\frac{2\Lambda}{16\pi G}\Big|^{-\frac{D}{2}}
  \int d^D x\;
  \sqrt{|\det R_{(\mu\nu)}(x)|}.
  \label{eq:edd}
\end{equation}
As in the original purely affine literature, this action is classically equivalent to GR in $D>2$ under the conditions listed above: we have only used the metric equations of motion algebraically, without introducing new dynamics.

The dependence on $V(\phi)$ (or on the cosmological constant in the pure gravity case) is central.  
For the metric to be well defined along the lines of \eqref{eq:metric-from-ricci}--\eqref{eq:edd-matter}, the potential must not cross zero in the region of field space explored by the theory.  
If one writes $V(\phi)=\frac{2\Lambda}{16 \pi G}+V_{\text{int}}(\phi)$ with an interaction $V_{\text{int}}(\phi)\geq 0$, then a negative cosmological constant $\Lambda<0$ generically implies the existence of field values for which $V(\phi)=0$, where the expression for $g_{\mu\nu}$ diverges. This signals that the purely affine description in terms of a regular emergent metric breaks down when $V(\phi)$ changes sign, suggesting that a positive cosmological constant is the most interesting parameter range to explore in this approach.

For our purposes, the crucial feature of \eqref{eq:edd} is that for a positive cosmological constant $\Lambda>0$ the action has a definite sign and is bounded below: the overall prefactor in \eqref{eq:edd} is positive and the integrand involves $\sqrt{|\det R_{(\mu\nu)}(x)|}$, which is non--negative. 
This contrasts sharply with the usual Einstein--Hilbert action formulated in terms of the metric, whose unboundedness due to the conformal mode is a long-standing obstacle to defining the gravitational path integral.  

In the rest of this work, we take the purely affine action \eqref{eq:edd} as our starting point and treat the affine connection as the fundamental field in a statistical theory. In particular, in the next section, we discretize \eqref{eq:edd} on a hypercubic lattice and use it to define a Gibbs ensemble of affine connections with Boltzmann weight $\exp(-S_{\text{Edd}})$.

\section{Partition function on a hypercubic lattice}
\label{sec:lattice-ricci}

To define a non-perturbative path integral for the purely affine theory, we discretize the manifold where the connection is defined using a fixed, periodic, hypercubic lattice and we approximate the Eddington action \eqref{eq:edd} in this lattice.  
The lattice is introduced purely as an ultraviolet regulator: it explicitly breaks diffeomorphism invariance and is not dynamical, in contrast to Causal Dynamical Triangulations \cite{Ambjorn:2012jv,Loll:2019rdj}.  
The fundamental dynamical variables on the lattice are the affine connection components $\Gamma^\lambda{}_{\mu\nu}(x)$, treated as real fields on the sites of the lattice.

We work on a $D$--dimensional manifold $M$ and discretize a coordinate patch into a hypercubic lattice with spacing $\varepsilon$ and periodic boundary conditions.  
At each lattice site $x$ we identify a tangent-space basis with the coordinate basis aligned along the links of the lattice.  
All indices ($\mu,\nu,\dots$) refer to this fixed frame, and repeated indices are summed over unless explicitly stated otherwise.


Given an affine connection $\Gamma^\alpha{}_{\beta\mu}(x)$ on the lattice site $x$, we can approximate derivatives by finite differences. For any lattice field $f(x)$ (with arbitrary index structure) we define the central finite difference $(\Delta^c_\mu f)(x)
  \;\equiv\;
  \frac{f(x+\varepsilon\hat\mu)-f(x-\varepsilon\hat\mu)}{2\varepsilon}$
where $\hat\mu$ is the unit vector in direction $\mu$.  
Applying this to the connection components gives a straightforward lattice approximation of the continuum expression for the Riemann tensor
\begin{equation}
  R^{\text{lat}\,\alpha}{}_{\beta\mu\nu}(x)
  \;\equiv\;
  \Delta^c_\mu\Gamma^\alpha{}_{\beta\nu}(x)
  -\Delta^c_\nu\Gamma^\alpha{}_{\beta\mu}(x)
  + \Gamma^\alpha{}_{\kappa\mu}(x)\,\Gamma^\kappa{}_{\beta\nu}(x)
  - \Gamma^\alpha{}_{\kappa\nu}(x)\,\Gamma^\kappa{}_{\beta\mu}(x)\,,
  \label{eq:finitedifferenceriemannlattice}
\end{equation}
which reduces to the continuum formula in the limit $\varepsilon\to 0$.

The corresponding lattice Ricci tensor is obtained by contraction:
\begin{equation}
  R^{\text{lat}}_{\rho\nu}(x)
  \;\equiv\;
  R^{\text{lat}\,\alpha}{}_{\rho\alpha\nu}(x)\,,
  \label{eq:Ricci-lat}
\end{equation}
and when needed, we take its symmetric part $R^{\text{lat}}_{(\mu\nu)}(x)
  \;\equiv\;
  \frac{1}{2}\left(R^{\text{lat}}_{\mu\nu}(x)+R^{\text{lat}}_{\nu\mu}(x)\right)$.

This finite-difference Ricci tensor is the basic approximation used in the main text of this work.  
In Appendix~\ref{sec:wilsonloops-improved} we describe an alternative discretization in which the curvature is reconstructed from gravitational Wilson loops on plaquettes.  
Both discretizations are expected to define the same continuum theory, differing only by lattice artifacts.

\subsection{Lattice Eddington action and statistical partition function}
\label{subsec:lattice-eddington}

We now discretize the purely affine Eddington action \eqref{eq:edd} by replacing the continuum Ricci tensor by its lattice counterpart and the integral by a sum over sites.  
In the pure gravity case, with $\Lambda>0$, this gives
\begin{equation}
  S_{\text{Edd}}^{\text{lat}}(\Gamma)
  \;=\; \frac{(D-1)(D-2)^{\frac{D-2}{2}}}{(16\pi G)^{\frac{D}{2}}}\, \Big(\frac{2\Lambda}{16\pi G}\Big)^{\frac{2-D}{2}}\;
  \sum_{x}\varepsilon^D\,\sqrt{\,\big|\det R^{\text{lat}}_{(\mu\nu)}(x)\big|\,}\,.
  \label{eq:lattice-edd}
\end{equation}
At the classical level, this approximation has a continuum limit $\varepsilon\to 0$ that recovers the Eddington action,
\begin{equation}
  \lim_{\varepsilon\to 0} S_{\text{Edd}}^{\text{lat}}(\Gamma) \;=\; S_{\text{Edd}}(\Gamma).
\end{equation}

Because the continuum action has a definite sign for $\Lambda>0$, the lattice action \eqref{eq:lattice-edd} is non--negative configuration by configuration. Furthermore, since for large values of the connection we have that (schematically)
\begin{equation}
    \lim_{\Gamma\to\infty}\sqrt{\,\big|\det R^{\text{lat}}_{(\mu\nu)}(x)\big|\,} \propto \big|\Gamma^2\big|^{\frac{D}{2}}\,,
\end{equation}
taming large values with at least Gaussian tails, this encourages us to define a statistical partition function in complete analogy with ordinary lattice field theory,
\begin{equation}
  Z_{\text{stat}}
  \;=\;
  \int \mathcal{D}\Gamma\;
  \exp\!\big[-S_{\text{Edd}}^{\text{lat}}(\Gamma)\big]\,,
  \label{eq:stat-partition}
\end{equation}
where $\mathcal{D}\Gamma$ denotes an appropriate product of Lebesgue measures over the connection components at each lattice site $\mathcal{D}\Gamma=\prod_{x,\lambda,\mu,\nu}\! d\Gamma^{\lambda}{}_{\mu\nu}(x)\times\text{(Gauge/Constraints)}$, subject to some constraints for different ensembles, to be specified below.  

This construction is deliberately modest: we simply treat $S_{\text{Edd}}^{\text{lat}}$ as an energy functional in a classical statistical mechanics model and interpret \eqref{eq:stat-partition} as a Gibbs measure over affine connections.  
We do \emph{not} derive this weight from a concrete Wick rotation of a Lorentzian path integral, nor do we assume that such a Wick rotation necessarily exists in the usual sense.  
As discussed in the introduction, this purely statistical point of view is common in non-perturbative approaches to quantum gravity, as it is out of the scope of this work to go beyond that. 

\subsection{Ensembles and path-integral measure}
\label{subsec:ensembles-measure}

The path integral \eqref{eq:stat-partition} is an integral over the affine connection at each lattice site.  
There are two natural choices for the integration domain:
\begin{itemize}
  \item a \emph{torsionful} ensemble, in which all $D^3$ components $\Gamma^\lambda{}_{\mu\nu}(x)$ are integrated over;
  \item a \emph{torsionless} ensemble, in which one restricts from the outset to symmetric connections $\Gamma^\lambda{}_{(\mu\nu)}(x)$ and enforces $\Gamma^\lambda{}_{[\mu\nu]}(x)=0$ at all sites.
\end{itemize}
These correspond to two different choices of configuration space for the statistical system; in both cases the connection is the only dynamical field.

Independent of this choice, the lattice already breaks continuum diffeomorphism invariance.  
However, there remain internal redundancies associated with the connection field itself.  
In the torsionful case there is a projective symmetry; in the torsionless case one must account for the Jacobian associated with restricting to symmetric connections.  
We now discuss these issues in turn and show that in both cases the corresponding Jacobians are field--independent, so no non--trivial ghost determinants arise.

\subsubsection{Torsionful ensemble}
\label{subsubsec:torsionful}

In the torsionful ensemble, the affine connection is not assumed to be symmetric in its lower indices.  
In this case, the symmetric part of the Ricci tensor is invariant under the projective transformation \cite{Garcia-Parrado:2020lpt}
\begin{equation}
    \Gamma^\alpha{}_{\beta \mu}(x) \;\longrightarrow\; \Gamma^\alpha{}_{\beta \mu}(x) + \delta^\alpha{}_{\beta}\,\eta_{\mu}(x)\,,
    \label{eq:gaugetransformationtorsionful}
\end{equation}
where $\eta_{\mu}(x)$ is an arbitrary vector field. This is a local gauge redundancy of the purely affine theory \cite{Ferraris:1981,Poplawski:2007} that must be fixed in the path integral.

A convenient gauge choice is
\begin{equation}
    \Gamma^\alpha{}_{\alpha \mu}(x)=0 \,,
    \label{eq:gaugefix-projective}
\end{equation}
which fixes the arbitrary projective degree of freedom at each site.  
The transformation \eqref{eq:gaugetransformationtorsionful} is a translation in field space,
\begin{equation}
\frac{\delta \Gamma'^{\alpha}{}_{\beta\mu}(x)}
{\delta \Gamma^{\rho}{}_{\sigma\nu}(y)}
= \delta^{\alpha}{}_{\rho}\,
  \delta^{\sigma}{}_{\beta}\,
  \delta^{\nu}{}_{\mu}\,
  \delta(x,y),
\end{equation}
so the Jacobian of the transformation in the measure $\mathcal{D}\Gamma$ is unity.
The associated Faddeev--Popov determinant for the gauge condition \eqref{eq:gaugefix-projective} is also a field--independent constant and can be absorbed into the normalization of $Z$.

One may therefore write the torsionful partition function either with an explicit gauge constraint,
\begin{equation}
Z_{\text{tf}}
\;=\;
\int \prod_{x,\lambda,\mu,\nu}\! d\Gamma^{\lambda}{}_{\mu\nu}(x)\;
\prod_{x,\sigma}\delta\!\big(\Gamma^{\rho}{}_{\rho\sigma}(x)\big)\;
e^{\!-\,S_{\text{Edd}}^{\text{lat}}(\Gamma)}\,,
\label{eq:integrationdeltaconstraintgaugefix}
\end{equation}
or, equivalently, with a Gaussian gauge--fixing term in the action,
\begin{equation}
Z_{\text{tf}}(\alpha_{\text{gf}})
\;=\;N(\alpha_{\text{gf}})
\int \prod_{x,\lambda,\mu,\nu}\! d\Gamma^{\lambda}{}_{\mu\nu}(x)\;
\exp\!\Big[-S_{\text{Edd}}^{\text{lat}}(\Gamma)
-\sum_{x,\sigma}\frac{\big(\Gamma^{\rho}{}_{\rho\sigma}(x)\big)^{2}}{2 \,\alpha_{\text{gf}}}\Big],
\label{eq:gaugefixinintegration}
\end{equation}
where $\alpha_{\text{gf}}$ is a gauge--fixing parameter and $N(\alpha_{\text{gf}})$ is an irrelevant normalization constant.  
Both forms are equivalent at the level of expectation values. In practice, one can either work with a strict traceless constraint such as \eqref{eq:integrationdeltaconstraintgaugefix}, or work directly with \eqref{eq:gaugefixinintegration}, letting the gauge--fixing term suppress excursions away from \eqref{eq:gaugefix-projective}.

\subsubsection{Torsionless ensemble}
\label{subsubsec:torsionless}

In the torsionless ensemble, the lower indices of the connection are constrained to be symmetric from the outset:
\begin{equation}
  \Gamma^\alpha{}_{\beta\mu}(x) = \Gamma^\alpha{}_{\mu\beta}(x)\,.
\end{equation}
There is no residual projective symmetry of the form \eqref{eq:gaugetransformationtorsionful} within this restricted space of connections, since such transformations remove the connection from the torsionless constraint.  
However, one must check the Jacobian associated with restricting the functional integration to the symmetric sector.

A convenient way to see that this Jacobian is constant is to change variables from the full connection to its symmetric and antisymmetric parts,
\begin{equation}
S^{\alpha}{}_{\beta\mu} \equiv \Gamma^{\alpha}{}_{(\beta\mu)}, 
\qquad
A^{\alpha}{}_{\beta\mu} \equiv \Gamma^{\alpha}{}_{[\beta\mu]} .
\end{equation}
For each fixed index $\alpha$ and each unordered pair $\{\beta,\mu\}$, the transformation to these new variables has constant determinant
\begin{equation}
\begin{pmatrix}
S^{\alpha}{}_{\beta\mu} \\[4pt]
A^{\alpha}{}_{\beta\mu}
\end{pmatrix}
= \frac{1}{2}
\begin{pmatrix}
1 & 1 \\[4pt]
1 & -1
\end{pmatrix}
\begin{pmatrix}
\Gamma^{\alpha}{}_{\beta\mu} \\[4pt]
\Gamma^{\alpha}{}_{\mu\beta}
\end{pmatrix}\qquad\to\qquad \det\!\left(\frac{1}{2}
\begin{pmatrix}
1 & 1 \\[4pt]
1 & -1
\end{pmatrix}\right)
= -\frac{1}{2}\,,
\end{equation}
so the Jacobian of the change of variables is a constant factor that can be absorbed into the overall normalization of the path integral.

Restricting to $A^{\alpha}{}_{\beta\mu}=0$ is then equivalent to inserting a product of delta functions in the measure. The torsionless partition function can be written as
\begin{equation}
Z_{\text{tl}}
\;=\;
\int \prod_{x,\lambda}\! \prod_{\mu\leq\nu}\!d\Gamma^{\lambda}{}_{(\mu\nu)}(x)\;
\prod_{x,\lambda}\prod_{\beta <\gamma}\!d\Gamma^{\lambda}{}_{[\beta\gamma]}(x)\,
\prod_{x,\lambda,\beta<\gamma}\delta\!\big(\Gamma^{\lambda}{}_{[\beta\gamma]}(x)\big)\;
e^{\!-\,S_{\text{Edd}}^{\text{lat}}(\Gamma)}\,.
\label{eq:integrationtorsionless}
\end{equation}
Here $\prod_{\mu\leq\nu}$ avoids double counting of symmetric pairs.  
Equivalently, one can integrate only over the symmetric components and drop the delta functions, since the antisymmetric components are fixed to zero.

In the numerical implementation described in Sec.~\ref{sec:numerics} we will show the numerical results of working with the torsionless ensemble, enforcing the symmetry $\Gamma^\lambda{}_{\mu\nu}=\Gamma^\lambda{}_{\nu\mu}$ in each local update.  
The torsionful ensemble and its projective gauge fixing are discussed here mainly to show that both natural choices of configuration space lead to simple, local measures with trivial Jacobians.

\subsection{Size of theory space and comparison to metric variables}
\label{subsec:theoryspace}

It is useful to quantify the size of the configuration space of the purely affine theory and compare it with the familiar metric formulation.

In $D$ dimensions, a general affine connection $\Gamma^\lambda{}_{\mu\nu}$ has $D^3$ real components at each spacetime point.  
In the torsionless ensemble we impose symmetry in the lower indices, $\Gamma^\lambda{}_{\mu\nu} = \Gamma^\lambda{}_{\nu\mu}$, so the number of independent components per point is
\begin{equation}
  N_{\Gamma}^{\text{(torsionless)}} = D \times \frac{D(D+1)}{2} = \frac{D^2(D+1)}{2}.
\end{equation}
In the torsionful ensemble, we keep all $D^3$ components, but as mentioned before, there is a projective gauge symmetry which, when fixed, removes $D$ gauge directions per point, leaving
\begin{equation}
  N_{\Gamma}^{\text{(torsionful, gauge-fixed)}} = D^3 - D
\end{equation}
independent components per point after projective gauge fixing.

For comparison, the metric tensor $g_{\mu\nu}$ has $N_g = D(D+1)/2$ components.  
There is a substantial difference in the number of components to be integrated over in both cases.

This shows that, at fixed dimension $D$, the purely affine theory requires the exploration of a much larger local theory space than the metric formulation.  
In four dimensions, for example, the metric has ten components per point, while the torsionless affine connection has forty, and the torsionful, projectively gauge-fixed connection has sixty.  
This increase in the number of local integration variables is the main price to pay for taking the connection as fundamental and avoiding coordinate gauge fixing at the microscopic level, and it is one of the reasons why numerical simulations in the affine formulation are computationally expensive. The tradeoff of the affine approach is the boundedness of the action, which allows a path integral formulation, as discussed before.

\subsection{Lattice symmetries}
\label{subsec:lattsymmetries}
While the periodic hypercubic lattice breaks diffeomorphism invariance, there are still symmetries of the discretized system that we do not fix. The periodic lattice admits two discrete symmetries: $90$-degree rotations and translations along lattice edges. Since the lattice is finite and periodic, both of these symmetries have a finite volume and leave both the integration measure and the discretized action invariant. Therefore, not fixing them cannot lead to divergences or runaway behaviour in the simulations.

At the same time, the existence of these symmetries implies that no lattice vertex or lattice direction is physically distinguished. In particular, local quantities evaluated at a specific vertex label (or in a specific coordinate direction) are not, by themselves, symmetry-invariant observables; their expectation values must be independent of the chosen vertex label and direction. Equivalently, one may promote such local quantities to symmetry-invariant observables by averaging them over the lattice within each configuration.

 We will come back to the definition of particular observables later on.

\section{Computational implementation}
\label{sec:numerics}

In this section we describe the numerical implementation of the lattice theory defined above.  
We have developed a C++ code \cite{silva_lqagmc_2025} that simulates the purely affine Eddington action on a hypercubic lattice for different choices of ensemble and curvature discretization, in arbitrary dimension $D \ge 2$.  
All numerical results shown in this work are obtained with this code.

\subsection{Ensembles and discretizations}

The code is capable of realizing both statistical ensembles introduced in Sec.~\ref{subsec:ensembles-measure}, defined by the lattice Eddington action $S_{\text{Edd}}^{\text{lat}}(\Gamma)$ and the Boltzmann weight $\exp[-S_{\text{Edd}}^{\text{lat}}]$:
\begin{itemize}
  \item A \emph{torsionful ensemble}, in which all components $\Gamma^\lambda{}_{\mu\nu}(x)$ are integrated over, subject only to the projective gauge condition \eqref{eq:gaugefix-projective}.  
        The lower indices of the connection are not constrained to be symmetric.
  \item A \emph{torsionless ensemble}, in which one restricts from the outset to symmetric connections $\Gamma^\lambda{}_{(\mu\nu)}(x)$ and enforces $\Gamma^\lambda{}_{[\mu\nu]}(x)=0$ at all sites, as in Eq.~\eqref{eq:integrationtorsionless}.  
        In this case there is no residual projective symmetry in the integration domain.
\end{itemize}
These correspond to two different choices of integration domain for the affine theory, and we will loosely refer to them as two ``ensembles'' in what follows.  
In both cases, the fundamental dynamical variable is the affine connection; the metric, when it exists, is always a derived quantity.

Independent of this choice, the code offers two discretizations of the Ricci tensor that can be plugged into the same action:
\begin{itemize}
  \item A \emph{finite-difference discretization}, in which $R^{\text{lat}}_{\mu\nu}(x)$ is computed from central finite differences of the affine connection, based on Eqs.~\eqref{eq:finitedifferenceriemannlattice}.
  \item A \emph{Wilson-loop discretization}, in which an approximate Ricci tensor $\tilde{R}^{\text{lat}}_{\mu\nu}(x)$ is reconstructed from gravitational Wilson loops on plaquettes, as described in Appendix~\ref{sec:wilsonloops-improved}.
\end{itemize}
These should be viewed as two regularizations of the same continuum observable; they define the same ensemble in the continuum limit, up to discretization effects. In our simulations, we checked that the obtained results do not depend on this choice, although using the Wilson loops is approximately an order of magnitude slower, due to the larger number of computations needed in each of the updates described below.

\emph{For all numerical results presented in this paper we use the torsionless ensemble together with the finite-difference Ricci tensor.}  
The torsionful ensemble and the Wilson-loop discretization are implemented in the code and can be used for further studies, but they are not employed in the production runs reported below.

\subsection{Local Metropolis updates of the connection}

As mentioned before, we work on a periodic hypercubic lattice, with linear extents $(N_1,\dots,N_D)$, where each $N_i$ counts the number of lattice sites in each dimension, and spacing $\varepsilon$, so that the total number of lattice sites is $N=\prod_\mu N_\mu$. 
On each site, we store all components of the connection $\Gamma^\lambda{}_{\mu\nu}(x)$ in a flat array, together with auxiliary fields holding the current Ricci tensor $R^{\text{lat}}_{\mu\nu}(x)$ and its determinant.

Configuration updates are generated by a standard Metropolis algorithm acting locally on a single lattice site $x$ at a time.  
One Monte Carlo sweep consists of visiting all lattice sites once (in a fixed or random order) and, at each site, performing the following steps:
\begin{enumerate}
  \item \emph{Identify the affected region.}  
  Changing $\Gamma^\lambda{}_{\mu\nu}(x)$ only affects the curvature in a finite neighbourhood of $x$, determined by the finite-difference stencil (or by the set of plaquettes touching $x$ in the Wilson-loop discretization).  
  We precompute the list of lattice sites whose Ricci tensor must be recomputed when $\Gamma$ at $x$ is changed.

  \item \emph{Compute the old local action.}  
  For all affected sites we read the current Ricci tensors $R^{\text{lat}}_{\mu\nu}$ and compute their contributions to $S_{\text{Edd}}^{\text{lat}}$, summing over those sites only.  
  This defines a local ``old action'' $S_{\text{loc}}^{\text{old}}$.

  \item \emph{Gaussian proposal for the connection.}  
  At the chosen site $x$ we propose a new value of the connection by adding a Gaussian random perturbation to all components at once,
  \begin{equation}
    \Gamma^\lambda{}_{\mu\nu}(x) \;\longrightarrow\;
    \Gamma^\lambda{}_{\mu\nu}(x) + \delta\Gamma^\lambda{}_{\mu\nu}(x),
  \end{equation}
  where each $\delta\Gamma^\lambda{}_{\mu\nu}(x)$ is drawn independently from a normal distribution with zero mean and variance $\sigma^2$.  
  In the torsionless ensemble, the proposal is immediately projected onto the symmetric sector by enforcing $\Gamma^\lambda{}_{\mu\nu}(x)=\Gamma^\lambda{}_{\nu\mu}(x)$ after the update.  
  In the torsionful case, the code instead works with the full set of components and imposes the projective gauge condition \eqref{eq:gaugefix-projective}.

  \item \emph{Compute the new local action.}  
  Using the updated connection at $x$, we recompute the Ricci tensor on all affected sites and sum their contributions to obtain a new local action $S_{\text{loc}}^{\text{new}}$.

  \item \emph{Metropolis test.}  
  The proposed change is accepted with probability
  \begin{equation}
    P_{\text{acc}} = \min\bigl\{1,\exp[-(S_{\text{loc}}^{\text{new}} - S_{\text{loc}}^{\text{old}})]\bigr\},
  \end{equation}
  which is the usual Metropolis rule, guaranteeing detailed balance.  
  If the move is accepted, we keep the new $\Gamma^\lambda{}_{\mu\nu}(x)$ and overwrite the stored Ricci tensors in the affected region with their new values; otherwise, we restore the old connection at $x$ and discard the trial Ricci tensors.
\end{enumerate}

The step size $\sigma$ of the Gaussian proposal is tuned during the simulation to keep the acceptance rate in a reasonable range, more precisely between $30\%$ and $60\%$, using a simple feedback rule based on the measured acceptance in each sweep.  
Once a satisfactory acceptance rate is reached, $\sigma$ is kept fixed for the remainder of the run. If the acceptance rate drops below $30\%$, then $\sigma$ is reduced by a factor of $1.2$, and if the acceptance rate increases above $60\%$, then $\sigma$ is increased by a factor of $1.2$. At least in the test case analyzed later on in this manuscript, this proved to be an efficient mechanism to keep the acceptance rate after thermalization close to $35\%$. 
We checked that updating each component individually increases the acceptance rate, but also increases simulation time substantially, so we compromised in this local, more abrupt proposal for all components at once.
This local update scheme exploits the locality of the action and is sufficient for the exploratory simulations presented here; more sophisticated algorithms (such as hybrid Monte Carlo) could be implemented within the same framework in future studies.

\subsection{General analysis strategy and observables}

Ultimately, in dimensions $D>2$ one would like to characterize the continuum limit of the theory through observables. Observables should be invariant under the symmetries of the theory, and thus they must be diffeomorphism invariant. 

In the current ensembles, we do not constrain the connection to be such that it gives rise to an invertible Ricci tensor, nor such that it has a definite signature to define Euclidean distances. As stressed in Sec.~\ref{sec:palatini}, the emergent metric obtained from the affine variables is not automatically well defined on every configuration: as the formalism is now, the path integral integrates over all connections compatible with the chosen torsion and gauge conditions, without imposing further constraints on it. 
This means that we are not guaranteed that the Ricci tensor can define a meaningful metric in the sense given by \eqref{eq:metric-from-ricci}.

This unconstrained connection ensemble therefore prevents us from measuring geometric characteristics such as Hausdorff or spectral dimensions, in the spirit of other lattice approaches to quantum gravity \cite{Hamber:2009,Ambjorn:2012jv,Loll:2019rdj}. Nevertheless, there are other easily accessible observables whose expectation value can be computed in this setting.

A simple and easily accessible choice is to monitor local quantities constructed from contractions of the Riemann tensor, since this tensor is readily available in lattice simulations. In particular, constructing diffeomorphism-invariant quantities without a notion of a metric that allows for raising and lowering indices reduces the number of possibilities. Nevertheless, there are those that do not require raising and lowering of indices, and are easily accessible in this setting. The path taken here to build observables is to compute integrals of tensor densities. Luckily, in any dimension $D$, we can build volume measures from $(0,\,2)$ tensors built purely from the affine connection, as we will describe below. 

A canonical choice for such an observable that \emph{would be} proportional to the physical volume of the system \emph{when a metric can be defined} from \eqref{eq:metric-from-ricci}, is
\begin{equation}
    V_{R}\equiv \int d^{D}x \sqrt{|\det (R_{(\mu \nu)}(x))|},
    \label{eq:integratedriccidensity}
\end{equation}
where the integral over the manifold is replaced by a discrete sum in the lattice simulations. 

Another possibility, given the ensembles we work with, is to build another volume measure from a tensor called the homothetic tensor
\begin{equation}
   Q_{\mu \nu} \equiv R^{\alpha}_{\,\,\,\alpha \mu \nu}\,,
   \label{eq:homotheticcurvaturedef}
\end{equation}
an antisymmetric tensor, which is often called the ``curl of the connection''. This tensor is zero in Riemannian geometry, and colloquially, it is said to measure how much the affine connection fails to preserve volume under parallel transport, and shows quantitative deviations from Riemannian manifolds, see \cite{ponomarev2017gauge}. Because it is skew-symmetric, its determinant is always greater than or equal to zero, and from it, we can integrate the density and obtain
\begin{equation}
     V_{Q}\equiv \int d^{D}x \sqrt{\det (Q_{\mu \nu}(x))}\,,
     \label{eq:integratedQdensity}
\end{equation}
where again, the integral over the manifold is replaced by a discrete sum in the lattice simulations.

Notice that any quantity that is built from \eqref{eq:integratedriccidensity} and \eqref{eq:integratedQdensity} by just multiplying with a constant scalar is also an observable. In the lattice simulations, we will compute in each path integral configuration the quantities
\begin{equation}
   \overline{V^{\text{lat}}_{R}}\equiv \frac{1}{N}\sum_{x} \varepsilon^{D} \sqrt{|\det (R^{\text{lat}}_{(\mu \nu)}(x))|} \qquad \qquad \overline{V^{\text{lat}}_{Q}}\equiv \frac{1}{N}\sum_{x} \varepsilon^{D} \sqrt{\det (Q^{\text{lat}}_{\mu \nu}(x))}\,,
   \label{eq:observablesonlattice}
\end{equation}
with $N$ the number of lattice sites, and where $\overline{V^{\text{lat}}_{R}}$ and $\overline{V^{\text{lat}}_{Q}}$ could be intuitively interpreted as the average physical volume, in the sense that the Ricci is proportional to the emergent metric in suitable scenarios \eqref{eq:metric-from-ricci}, and the average volumetric deviation from Riemannian geometry of each hypercube of the lattice, respectively. 

It is worth highlighting that these observables are independent of the labeling of the vertices, as well as from particular lattice directions around a particular point, and are thus free from the problem of discrete lattice symmetries explained before. 

On top of the previously constructed observables, it is interesting to look at another order parameter. As previously mentioned, the ensembles are not constrained to give rise to symmetric Ricci tensors with fixed signature, and it is thus interesting to study whether the system has a particular preference for the signature of the resulting tensors in the simulations. In particular, an easily accessible quantity to look at is the fraction of points in the lattice that have $\det (R^{\text{lat}}_{(\mu \nu)}(x))<0$, meaning that at least one of its eigenvalues is negative
\begin{equation}
  f_{d_{R}<0} \equiv \frac{1}{N}||\{ x \, /\, det (R^{\text{lat}}_{(\mu \nu)}(x))<0\}\,||\,.
  \label{eq:fracnegativedeterminantsymricci}
\end{equation}
In the toy model studied in this work, where $D=2$, the fraction $f_{d_{R}<0}$ measures the percentage of points in the manifold that have one negative and one positive eigenvalue. Since changes in the signature imply a change of signature of one eigenvalue, and that therefore it is zero in some parts of the manifold, this fraction also provides information about how the symmetric Ricci tensor is non-invertible in some parts of the manifold. In a more constrained ensemble, one should integrate over connections that give rise to an invertible symmetric part of the Ricci tensor, providing a well-defined metric tensor.
 
In the two-dimensional study of Sec.~\ref{sec:2D-results} we use the observables \eqref{eq:observablesonlattice} and the order parameter \eqref{eq:fracnegativedeterminantsymricci} as thermalization indicators, and we compute their expectation values as a function of the bare coupling of Eddington's action. 

\subsection{Units}
The affine connection has units of inverse length in all dimensions. It is interesting to see that in the hypercubic lattice approximation of Eddington's action \eqref{eq:lattice-edd}, the uniform lattice spacing $\varepsilon$ can be absorbed in a new dimensionless connection variable 
\begin{equation}
    \tilde{\Gamma}^{\sigma}_{\,\mu \nu}\equiv\varepsilon\, \Gamma^{\sigma}_{\,\mu \nu}.
    \label{eq:dimlessconnection}
\end{equation}
The transformation to these dimensionless variables has a trivial path integral Jacobian, and will be used for simulations. Notice that when using these variables, the lattice spacing disappears from all the relevant quantities that are used in this work, obtained by integrating over the entire manifold, such as the action, or the observables \eqref{eq:observablesonlattice}. For other quantities that involve lower-dimensional integrals, such as some suitable measure of distance that will be discussed in the appendix, the lattice spacing appears again. Still, since in this work we will restrict the analysis to the simple quantities defined in \eqref{eq:observablesonlattice}, the lattice spacing will play no role. This transformation can effectively be thought of as setting $\varepsilon = 1$, which is what we will do in the following.


\section{Two-dimensional results}
\label{sec:2D-results}

As discussed in Sec.~\ref{sec:palatini}, in $D=2$ the metric cannot be reconstructed from the affine variables, but the purely affine action itself has a smooth and non-trivial $D\to 2$ limit. Besides, the two-dimensional case is the one with the lowest number of connection components, an appealing feature for testing simulation aspects. 
This makes two dimensions an attractive testing ground: the theory remains dynamically non-trivial due to its non-vanishing action, yet any direct comparison to metric formulations of gravity is impossible by construction.  
Before describing the numerical experiments, let us make this $D\to2$ limit explicit. 

\subsection{Action in the $D\to2^+$ limit}

In the pure gravity case, the continuum Eddington action in $D$ dimensions can be written as
\begin{equation}
  S_{\text{Edd}}[\,\Gamma\,]
  \;=\;
  \alpha_D(\Lambda,G)
  \int d^D x\,\sqrt{\big|\det R_{(\mu\nu)}(x)\big|}\,,
  \label{eq:SEdd-alphaD}
\end{equation}
with
\begin{equation}
  \alpha_D(\Lambda,G)
  \;\equiv\;
  \frac{(D-1)(D-2)^{\frac{D-2}{2}}\,2\Lambda}{(16\pi G)^{\frac{D+2}{2}}}
  \Big|\frac{2\Lambda}{16\pi G}\Big|^{-\frac{D}{2}}.
  \label{eq:alphaD-def}
\end{equation}

For definiteness, take $\Lambda>0$ so the absolute values can be dropped.  
The $(D-2)^{\frac{(D-2)}{2}}$ has a finite, real value when approaching $D\to 2^+$, thus, in two dimensions, the Eddington reduces to
\begin{equation}
  \boxed{%
  S_{\text{Edd}}^{(2D)}[\,\Gamma\,]
  \;=\;
  \alpha
  \int d^2 x\,\sqrt{\big|\det R_{(\mu\nu)}(x)\big|}\,,
  \qquad
  \alpha\equiv \alpha_2(\Lambda,G)= \frac{1}{16\pi G}.}
  \label{eq:SEdd-2D-continuum}
\end{equation}
The dependence on $\Lambda$ has dropped out completely in this limit; only $G$ survives in the overall prefactor.  

On the lattice, the continuum integral is replaced by a sum,
\begin{equation}
  S_{\text{Edd}}^{(2D)\,\text{lat}}[\,\Gamma\,]
  \;=\;
  \alpha
  \sum_x \varepsilon^2 \sqrt{\big|\det R^{\text{lat}}_{(\mu\nu)}(x)\big|}\,,
  \label{eq:SEdd-2D-lattice}
\end{equation}
with $R^{\text{lat}}_{(\mu\nu)}(x)$ given by the finite-difference discretization introduced in Sec.~\ref{sec:lattice-ricci} (or its Wilson-loop improvement).  
Strictly speaking, the local ``volume element'' entering the action is $\varepsilon^2\sqrt{|\det R^{\text{lat}}_{(\mu\nu)}(x)|}$; in the simulations reported below, we fix $\varepsilon=1$, basically absorbing the lattice spacing into the connection variables as described in \eqref{eq:dimlessconnection}. 

It is convenient to treat the prefactor $\alpha$ in front of the sum as a tunable bare coupling in the statistical ensemble, and we explore the statistical ensemble defined by the Boltzmann weight
\begin{equation}
  Z(\alpha) \;=\; \int \mathcal{D}\Gamma\,e^{- S_{\text{Edd}}^{(2D)\,\text{lat}}[\,\Gamma\,]}.
\end{equation}
Here $\alpha$ appears as a dimensionless parameter controlling how strongly configurations with large local curvature determinant are suppressed; in practice, we probe several values of $\alpha$ in a reasonable range, treating it as a bare coupling.

\subsection{Simulation setup in two dimensions}

While the provided code supports simulations in $D$ dimensions, and clearly the most physically interesting case would be the four-dimensional one, the computational cost scales at least with the size of the simulated manifold, which for $N$ lattice points in a given direction is equal to $N^{D}$. On top of that, at this stage, where we are interested in a proof of concept implementation, and where we do not constrain the ensemble of affine connections to generate fixed signature emergent metrics, we consider this exercise unnecessary, and we restrict the analysis in this work to the simplest possible case, $D=2$.

All simulations reported here are performed on periodic hypercubic lattices of sizes $N=50\times 50, \,N=100\times 100$ and $N=150\times 150$ in $D=2$, with $\varepsilon=1$.  
The dynamical variable is the affine connection $\Gamma^\lambda{}_{\mu\nu}(x)$ at each site, updated with the local Gaussian Metropolis algorithm described in Sec.~\ref{sec:numerics}.
The local Gaussian standard deviation is always initialized at $\sigma=1$, and once the system is thermalized it typically drops to around $\sigma\simeq0.12$ for the smaller values of $\alpha$ explored in this work, and $\sigma\simeq0.08$ for the larger values of $\alpha$ explored in this work, with an acceptance ratio that stabilizes between $33\%$ to $37\%$. 

We use the finite-difference discretization of the Ricci tensor and work in the torsionless ensemble, enforcing symmetry in the lower indices after each local update.

For each value of the coupling $\alpha$, we generate a Markov chain starting from the ``cold'' configuration
\begin{equation}
  \Gamma^\lambda{}_{\mu\nu}(x) = 0 \qquad \text{for all sites } x,
\end{equation}
projected onto the torsionless sector. We checked that starting from a ``hot'' configuration, where each component at each site was sampled from a Gaussian distribution, did not change the results. 
We then perform a large number of full-lattice sweeps.  
By studying the convergence of the observables \eqref{eq:observablesonlattice} and the order parameter \eqref{eq:fracnegativedeterminantsymricci}, we established that the first $10000$ to $15000$ sweeps can be regarded as thermalization, depending on the values of $\alpha$ used here. Therefore, for safety, we always discard the first $25\,000$ sweeps as thermalization, while the rest of the sweeps (up to $250\,000$ in total) are used for measurements.
Configurations are recorded every $100$ sweeps, allowing every vertex of the manifold to be updated $100$ times between measurements, a number of sweeps considered sufficient at this exploratory stage to reduce correlations between successive measurements, giving a total of $2000$ measured configurations for each value of $\alpha$. As a cross-check for this number of sweeps between measurements, we checked that measuring every $75$, $150$, and every $200$ sweeps gives the same results within error bars. The code exports the values of the connection at each site for each configuration, and has the option to export the discretized Ricci tensor as well.

\subsection{Numerical results}

In $D=2$ it is impossible to relate this affine theory to General relativity, since \eqref{eq:metric-from-ricci} breaks down and therefore no GR metric tensor exists.
However, as discussed before, we can still measure the observables \eqref{eq:observablesonlattice} that provide different notions of manifold volume and deviations from Riemannian geometry. 

The Monte Carlo data are processed as follows. At a fixed value of $\alpha$, for each saved configuration $c$ in the Markov chain, we first compute a certain observable or order parameter, generally denoted as $\mathcal{O}(c)$ on that configuration, and later approximate the path integral expectation value as
\begin{equation}
  \langle \mathcal{O} \rangle \;\approx\; \frac{1}{N_{\text{cfg}}}\sum_{c=1}^{N_{\text{cfg}}} \mathcal{O}(c)\,,
\end{equation}
where $c$ labels the saved configurations, and $N_{\text{cfg}}$ is the number of stored configurations at that value of $\alpha$. In this work, we used $N_{\text{cfg}}=2000$ for all the studied values of $\alpha$.

\begin{figure}
    \centering
    \includegraphics[width=0.7\linewidth]{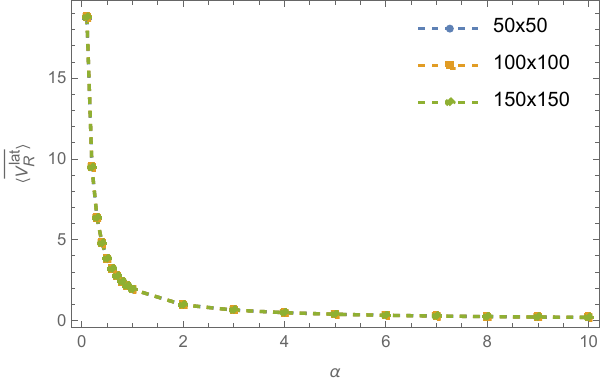}
    \caption{Expectation value of the observable $\overline{V^{\text{lat}}_{R}}$ over $2000$ sampled path integral configurations from the Monte Carlo simulations, as a function of the bare parameter $\alpha$, for three different lattice sizes. The results are overlapping monotonically decreasing curves, with no signs of a phase transition. Statistical error bars are below $0.01\%$ of the reported values, and are too small to be seen.}
    \label{fig:VolRicci}
\end{figure}

\begin{figure}
    \centering
    \includegraphics[width=0.7\linewidth]{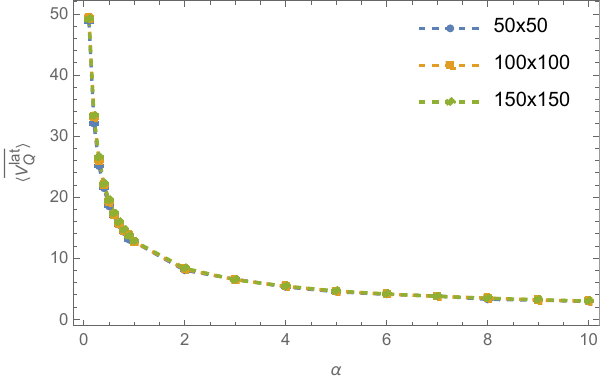}
    \caption{Expectation value of the observable $\overline{V^{\text{lat}}_{Q}}$ over $2000$ sampled path integral configurations from the Monte Carlo simulations, as a function of the bare parameter $\alpha$, for three different lattice sizes. The results are overlapping monotonically decreasing curves, with no signs of a phase transition. Statistical error bars are below $0.01\%$ of the reported values, and are too small to be seen.}
    \label{fig:VolQ}
\end{figure}

\begin{figure}
    \centering
    \includegraphics[width=0.7\linewidth]{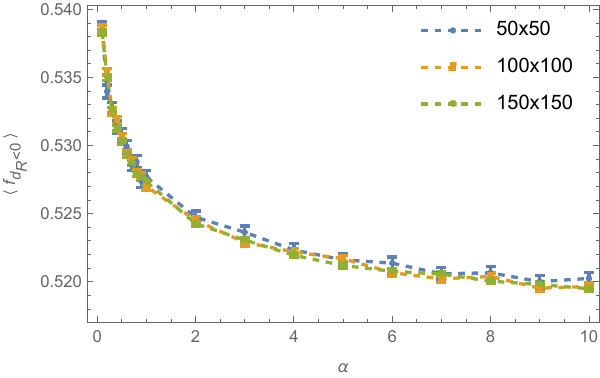}
    \caption{Expectation value of $f_{d_{R}<0}$ over $2000$ sampled path integral configurations from the Monte Carlo simulations, as a function of the bare parameter $\alpha$, for three different lattice sizes. The results are overlapping monotonically decreasing curves, with no signs of a phase transition, but always above $0.5$, meaning that the configurations have a majority of points with a symmetric part of the Ricci tensor with $(-,\,+)$ signature, instead of $(-,\,-)$ or $(+,\,+)$.}
    \label{fig:FracNegDetRicci}
\end{figure}

For each of the lattice sizes $N=50\times 50, \,N=100\times 100$ and $N=150\times 150$, we studied a range of $\alpha$ values that covers two orders of magnitude, specifically $\alpha \, \in \, \{0.1,\,0.2,\,\dots,\,0.9,\,1,\,2,\,\dots,\,10\}$.
Within this range, we computed the expectation value of the observables $\overline{V^{\text{lat}}_{R}}$ and $\overline{V^{\text{lat}}_{Q}}$ defined in \eqref{eq:observablesonlattice}, and the order parameter $f_{d_{R}<0}$ defined in \eqref{eq:fracnegativedeterminantsymricci}. The numerical results of each of them can be seen in Figs. \ref{fig:VolRicci}, \ref{fig:VolQ} and \ref{fig:FracNegDetRicci} respectively. 

Within the explored range of couplings, both observables $\overline{V^{\text{lat}}_{R}}$ and $\overline{V^{\text{lat}}_{Q}}$ and the order parameter $f_{d_{R}<0}$ vary smoothly and monotonically with $\alpha$. We do \emph{not} see any clear sign of non-monotonic behaviour or singular features that could be interpreted as a phase transition in terms of the measured quantities. 
If a non-trivial phase structure exists in the two-dimensional model, it does not show up through the measured geometrical properties on the lattices and coupling range used here.

There is, nevertheless, an interesting feature to be highlighted. Despite not having imposed a constraint on the ensemble of connections, such that there is a preferred signature of the symmetric part of the Ricci tensor, it seems that the system thermalizes on configurations that have approximately between $52\%$ and $54\%$ of their domain covered by a tensor $R_{(\mu \nu)}$ with signature $(-,\,+)$. It remains to be seen in future work if a more sophisticated ensemble, perhaps with further constraints on the affine connection, changes this small preference for signature average. 

All results of the observables \eqref{eq:observablesonlattice} and the order parameter \eqref{eq:fracnegativedeterminantsymricci} overlap within statistical error bars, meaning that we do not observe any non-trivial scaling within the explored range of lattice sizes, and the factor $\frac{1}{N}$ in the definition of each quantity is sufficient to overlap all curves.

Given the special role of two dimensions in the classical theory, the present results should be viewed as a proof of principle rather than as a test of continuum General Relativity.  
They show that the purely affine Eddington-type action leads to a well-defined lattice ensemble whose basic properties can be studied numerically, and that the coupling $\alpha$ provides a clean handle to scan through different regimes of the model.  
In higher dimensions, where the metric can be reconstructed, one can build more intuitive geometric observables (such as scaling dimensions) and use similar Monte Carlo techniques to investigate the continuum limit and possible phase structure of the theory.

The main lesson from these two-dimensional simulations is that the purely affine lattice theory defined by $S_{\text{Edd}}^{(2D)\,\text{lat}}$ behaves as a perfectly ordinary statistical model of a high-dimensional field, even though no metric can be reconstructed in $D=2$.  
The action prefactor $\alpha$ monotonically controls the typical scale of the diagnostic observables built from the Riemann tensor, and the simple local Gaussian Metropolis dynamics is able to explore the configuration space without apparent instabilities or pathologies.

\section{Summary and outlook}
\label{sec:concl}

In this work, we took the first steps towards developing and exploring a new non-perturbative lattice framework for the quantization of gravity, in which the affine connection is taken as the fundamental field and the metric, when it exists, is an emergent quantity. Starting from the Palatini formulation of general relativity with matter, we recalled how, in dimensions higher than two and under mild assumptions, the metric can be eliminated algebraically in favour of the symmetric Ricci tensor, leading to an Eddington-type purely affine action that is classically equivalent to general relativity. This action has a definite sign, being positive for a positive cosmological constant, making a statistical treatment natural using its action as a Boltzmann weight. On a fixed hypercubic lattice, we discretized this action and constructed both torsionful and torsionless ensembles.

We complemented this construction with an explicit Monte Carlo implementation. The accompanying C++ code simulates the purely affine lattice action in arbitrary dimension using a standard local Metropolis algorithm, and offers both torsionful and torsionless ensembles, as well as different possible discretizations of the Ricci tensor. In the present work, we focused our numerical analysis on the simplest non-trivial case, a two-dimensional torsionless ensemble with a finite-difference discretization of the Ricci tensor. Even though no emergent metric exists in two dimensions, the purely affine action has a smooth limit and can be used to define a statistical model of the connection. We explored the ensemble as a function of the only bare coupling of the theory, and monitored the expectation value of two diffeomorphism-invariant geometric quantities, related to volume measures constructed from the Riemann tensor. Their expectation values vary smoothly with the bare coupling, showing a monotonic decrease as a function of it. Our data does not show any phase transition, but shows that the model is numerically well-behaved, presenting no divergences in the explored ranges, nor runaway behaviors.

The developed framework and the simulation code open up several directions for further work.

On the conceptual side, it will be important to understand how to treat configurations that do not admit a regular emergent metric, whether and how to restrict or reweight the ensemble to favour metric-compatible configurations, and how to define appropriate observables in a formulation where the connection is fundamental. We have not imposed configuration-by-configuration conditions ensuring the existence of a regular emergent metric, even in dimensions where this is possible in principle. It remains to be seen if a more constrained ensemble of affine connections that give rise to meaningful emergent metrics, via the mechanisms described in the first sections, can be constructed. For example, it would be interesting to integrate only over connections that give rise to ADM foliated symmetric parts of the Ricci tensor, to obtain a notion of time in each configuration, in a similar spirit to what is done in Causal Dynamical Triangulations. While it is not clear how such a constrained ensemble should be constructed, the number of components to be fixed is certainly large enough to satisfy both invertibility and a fixed signature of the symmetric part of the Ricci tensor. As it stands, the formulation is entirely statistical; we have not addressed the question of how, or whether, a Lorentzian, unitary quantum theory of gravity can be recovered from this framework. These are significant open issues that must be addressed before any firm claims about continuum quantum gravity can be made.

On the numerical side, once a suitable refined ensemble is constructed, extending the simulations to higher dimensions, in particular to four dimensions, would allow one to study genuinely geometric observables such as physical volumes, curvature correlators, effective scaling dimensions, and possibly the recovery of diffeomorphism invariance in the continuum limit. Improved update algorithms and curvature discretizations could help control lattice artifacts, making it possible to search for scaling regimes and fixed points in the spirit of other non-perturbative approaches. Moreover, a systematic continuum-limit existence analysis, involving several lattice spacings and volumes, has not yet been carried out. In higher dimensions, where the cosmological constant does not disappear from the action as it does in two dimensions, this analysis will be crucial for the determination of phase transitions that could define a continuum limit of this approach. 

 In a more refined ensemble, if a suitable Lorentzian continuation exists, as well as an appropriate phase transition of the lattice simulations, the quanta of this purely affine theory would be excitations of the connection rather than of the metric, suggesting a ``connecton'' picture of gravitational degrees of freedom, closer in spirit to the other fundamental interactions. Whether this speculative picture can be made precise, and whether the lattice framework developed here can serve as a good starting point for such a construction, are questions we will further explore in future work. 
 
 So far, we have shown that purely affine gravity, long regarded mainly as a classical reformulation, can be brought into the non-perturbative lattice arena in a concrete and computable way, and that its basic statistical properties can be explored with standard Monte Carlo techniques. We believe that this work is a first step into a rich research line in the quest for the long-standing problem of the quantization of gravity.

\section*{Acknowledgments}
I would like to thank Jan Ambjørn for extended and useful conversations we had throughout the development of these ideas. I would also like to thank Renate Loll, Jesse van der Duin, Marc Schiffer, Tom O'Shea, and Juan Fernandez Molinero for useful comments and feedback on this manuscript.

\clearpage

\section{Appendix}
\label{sec:technicalappendix}
\subsection{Physical lengths}
While the constructed hypercubic lattice has coordinate length $\varepsilon$, the physical length of an elementary link of the lattice $l=(x,\,x+\varepsilon\hat{\mu})$ with extremes $x$ and $x+\varepsilon\hat{\mu}$ is, in the purely affine formulation, supposed to be given by the emergent metric constructed from the symmetric part of the Ricci tensor as discussed in Sec.~\ref{sec:intro}. Whenever the algebraic relation between the metric and the symmetric Ricci tensor is well defined and yields an invertible tensor with the desired signature, one can assign a physical length and volume to the lattice elements.
For example, the square length of an elementary link of the lattice $l=(x,\,x+\varepsilon\hat{\mu})$ can be defined as (no sum over $\mu$)
\begin{equation}
    L^2_{phys}(l=(x,\,x+\varepsilon\hat{\mu}))=\varepsilon^2\, \left(\frac{D-2}{2 \Lambda}\right)\,\frac{1}{2}\,\left(  R^{lat}_{(\mu\mu)}(x) + R^{lat}_{(\mu\mu)}(x+\varepsilon\hat{\mu}) \right)\,,
\end{equation}
where a mid-point approximation was used. From this length, one could construct geodesic distances, and from it, covariant correlators, effective scaling dimensions, and many other geometrical properties.

If the metric is well defined, then one could also, for example, compute the total physical volume of the lattice using this induced metric,
\begin{equation}
    V_{phys}=|\frac{D-2}{2 \Lambda}|^{\frac{D}{2}}  \sum_{x}\varepsilon^{D}\sqrt{\big|\det R^{\text{lat}}_{(\mu\nu)}(x)\big|}\,.
\end{equation}

In particular, a Renormalization Group step in this setting could be done by increasing the lattice size and decreasing the lattice spacing, while simultaneously keeping fixed the physical volume $V_{phys}$, or other physical lengths. This was not explored in this work, but must be done in future work to explore the continuum limit of this approach.

\subsection{Ricci curvature from Wilson loops}
\label{sec:wilsonloops-improved}

For completeness, we now describe a more geometric discretization of the curvature, based on parallel transport around plaquettes (gravitational Wilson loops \cite{Wilson:1974sk}). This construction is closer in spirit to lattice gauge theory and provides an approximation to the continuum curvature that is manifestly built from holonomies. Its main drawback is that it requires repeated matrix exponentiations or accurate approximations thereof.


Consider an elementary square of the lattice in the domain of a chart of $M$, of side $\varepsilon$ in the $(\mu,\nu)$ plane, based at site $x$.
\begin{figure}[!ht]
    \centering
    \includegraphics[width=0.45\linewidth]{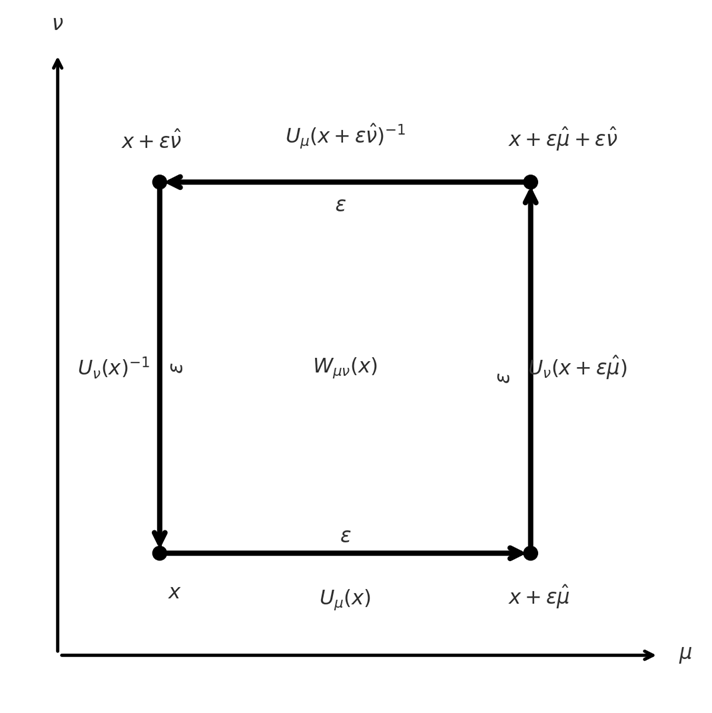}
    \caption{Wilson loop on a square plaquette with edge length $\varepsilon$.}
    \label{fig:placeholder}
\end{figure}

On this elementary square, or ``plaquette'' from now on, we will consider the parallel transport operator around a counterclockwise loop, as depicted in Fig. \ref{fig:placeholder}.  
Let $U_\mu(x)$ be the operator that transports vectors along $x\to x+\varepsilon\hat\mu$:
\begin{equation}
  V^\alpha(x+\varepsilon\hat\mu) \;=\; \big(U_\mu(x)\big)^\alpha{}_\beta\,V^\beta(x),
  \qquad
  U_\mu(x)=\mathcal{P}\exp\!\left(-\!\!\int_0^\varepsilon\!\!ds\,\Bar{\Gamma}_\mu(x+s\hat\mu)\right).
\end{equation}
where $\Gamma^\alpha{}_{\beta\mu}(x)$ is the affine connection and we defined the matrix $(\Bar{\Gamma}_\mu(x))^\alpha{}_\beta \equiv \Gamma^\alpha{}_{\beta\mu}(x)$ be the connection matrix.
To set an example of how one gets curvature from these variables, expanding to $\mathcal{O}(\varepsilon^2)$, one gets for the first two elements of the loop (here $\mu$ and $\nu$ are not summed over)
\begin{align}
  (U_\mu(x))^\lambda{}_{\rho} &= \delta^\lambda{}_{\rho}
   - \varepsilon\,\Gamma^\lambda{}_{\rho\mu}(x)
   + \frac{\varepsilon^2}{2}\Big(\partial_\mu\Gamma^\lambda{}_{\rho\mu}(x)
      + \Gamma^\lambda{}_{\kappa\mu}(x)\,\Gamma^\kappa{}_{\rho\mu}(x)\Big)
   + \mathcal{O}(\varepsilon^3),\\
  (U_\nu(x+\varepsilon\hat\mu))^\lambda{}_{\gamma} &= \delta^\lambda{}_{\gamma}
   - \varepsilon\Big(\Gamma^\lambda{}_{\gamma\nu}(x)+\varepsilon\,\partial_\mu\Gamma^\lambda{}_{\gamma\nu}(x)\Big)
   + \frac{\varepsilon^2}{2}\Big(\partial_\nu\Gamma^\lambda{}_{\gamma\nu}(x)+\Gamma^\lambda{}_{\beta\nu}\Gamma^\beta{}_{\gamma\nu}\Big)
   + \mathcal{O}(\varepsilon^3),
\end{align}
and similarly for the inverse matrices on the paths on the top and left edges of the loop in Fig. \ref{fig:placeholder}
\begin{equation}
(U_\mu^{-1})^{\lambda}{}_{\rho}
  = \delta^{\lambda}{}_{\rho}
    + \varepsilon\,\Gamma^{\lambda}{}_{\rho\mu}
    + \frac{\varepsilon^{2}}{2}\!\left(
        \Gamma^{\lambda}{}_{\kappa\mu}\,\Gamma^{\kappa}{}_{\rho\mu}
        - \partial_{\mu}\Gamma^{\lambda}{}_{\rho\mu}
      \right)
    + \mathcal{O}(\varepsilon^{3}) \, .
\end{equation}

From the oriented plaquette loop described in Fig. \ref{fig:placeholder}, we define the gravitational Wilson loop operator (here $\mu$ and $\nu$ are not summed over)
\begin{equation}
  (W_{\mu\nu}(x))^\alpha{}_{\beta}\;=\;(U_\mu(x))^\alpha{}_{\kappa}\,(U_\nu(x+\varepsilon\hat\mu))^\kappa{}_\delta\,(U_\mu^{-1}(x+\varepsilon\hat\nu))^\delta{}_\phi\,(U_\nu^{-1}(x))^\phi{}_\beta.
\end{equation}
Multiplying the four expanded links and keeping all $\mathcal{O}(\varepsilon^2)$ terms, the $\mathcal{O}(\varepsilon)$ pieces cancel pairwise and one obtains
\begin{align}
  \big(W_{\mu\nu}(x)\big)^\alpha{}_{\beta}
  &= \delta^\alpha{}_{\beta}
    - \varepsilon^2\Big(\partial_\mu\Gamma^\alpha{}_{\beta\nu}
                        -\partial_\nu\Gamma^\alpha{}_{\beta\mu}
                        + \Gamma^\alpha{}_{\kappa\mu}\Gamma^\kappa{}_{\beta\nu}
                        - \Gamma^\alpha{}_{\kappa\nu}\Gamma^\kappa{}_{\beta\mu}\Big)
    + \mathcal{O}(\varepsilon^3)\\[1mm]
  \big(W_{\mu\nu}(x)\big)^\alpha{}_{\beta}&= \delta^\alpha{}_{\beta} \;-\; \varepsilon^2\,R^\alpha{}_{\beta\mu\nu}(x) \;+\; \mathcal{O}(\varepsilon^3),
\end{align} 
where the Riemann tensor was identified in the last line.
Denoting the inverse of the Wilson loop $W_{\mu\nu}(x)$ as $W^{-1}_{\mu\nu}(x)=W_{\nu\mu}(x)$, since it is just the reverse direction loop, we get the expression at the lowest order in $\varepsilon$
\begin{equation}
  \boxed{\,R^\alpha{}_{\beta\mu\nu}(x) \;\simeq\; -\,\frac{1}{2\,\varepsilon^2}\big(\,W_{\mu\nu}(x)-W_{\nu\mu}(x)\,\big)^\alpha{}_{\beta}\;+\;\mathcal{O}(\varepsilon)\,}
\end{equation}
Which is the hypercubic lattice approximation of the curvature tensor on a plaquette $(\mu,\nu)$. A more sophisticated approximation could use a clover average of the four plaquettes touching $x$ in the $(\mu,\nu)$ plane, but for now this approximation suffices.

We are now in a situation where we can compute the Ricci tensor from Wilson loops. The previous setup of an hypercubic lattice allows us to reconstruct the Ricci tensor from the Wilson loops in an approximation up to the lowest order in the plaquette size
\begin{equation}
  \,R_{\rho\nu}(x) \;=\; R^\alpha{}_{\rho\alpha\nu} \longrightarrow \boxed{\,R^{\text{lat,W}}_{\rho\nu}(x) 
  \;\simeq\; -\frac{1}{2\,\varepsilon^2}\sum_{\alpha}\big(\,W_{\alpha\nu}(x)-W_{\nu\alpha}(x)\,\big)^{\alpha}{}_{\rho}\,}.
  \label{eq:Ricci-sum-Wilson}
\end{equation}

\paragraph{Approximation schemes for the Wilson-loop estimator. }
Since exponentiation of matrices to construct the link variables is computationally expensive, we will resort to a simple approximation of each link variable. We define ($\mu$ is not summed over)
\begin{align}
  (\Tilde{U}_\mu(x))^\lambda{}_{\rho} &\equiv \delta^\lambda{}_{\rho}
   - \varepsilon\,\Gamma^\lambda{}_{\rho\mu}(x)
   + \frac{\varepsilon^2}{2}\Big(\Gamma^\lambda{}_{\kappa\mu}(x)\,\Gamma^\kappa{}_{\rho\mu}(x)+\frac{1}{\varepsilon}\left(\Gamma^{\lambda}{}_{\rho\mu}(x+\varepsilon \hat{\mu})-\Gamma^{\lambda}{}_{\rho\mu}(x)\right)
      \Big),\\
  (\Tilde{U}_\mu^{-1}(x))^{\lambda}{}_{\rho}
  &\equiv  \delta^{\lambda}{}_{\rho}
    + \varepsilon\,\Gamma^{\lambda}{}_{\rho\mu}(x)
    + \frac{\varepsilon^{2}}{2}\!\Big(
        \Gamma^{\lambda}{}_{\kappa\mu}(x)\,\Gamma^{\kappa}{}_{\rho\mu}(x)
        - \frac{1}{\varepsilon}\left(\Gamma^{\lambda}{}_{\rho\mu}(x+\varepsilon \hat{\mu})-\Gamma^{\lambda}{}_{\rho\mu}(x)\right)
      \Big) \, .
\end{align}
which are accurate expressions for the link variables up to order $\mathcal{O}(\varepsilon^3)$. With these variables, we define the approximated Wilson loop 
\begin{equation}
  (\Tilde{W}_{\mu\nu}(x))^\alpha{}_{\beta}\;\equiv\;(\Tilde{U}_\mu(x))^\alpha{}_{\kappa}\,(\Tilde{U}_\nu(x+\varepsilon\hat\mu))^\kappa{}_\delta\,(\Tilde{U}_\mu^{-1}(x+\varepsilon\hat\nu))^\delta{}_\phi\,(\Tilde{U}_\nu^{-1}(x))^\phi{}_\beta.
\end{equation}
and the approximated lattice Ricci tensor
\begin{equation}
  \boxed{\,\Tilde{R}^{\text{lat,W}}_{\rho\nu}(x) 
  \;\simeq\; -\frac{1}{2\,\varepsilon^2}\sum_{\alpha}\big(\,\Tilde{W}_{\alpha\nu}(x)-\Tilde{W}_{\nu\alpha}(x)\,\big)^{\alpha}{}_{\rho}\,}.
  \label{eq:Ricci-sum-approx-lattice}
\end{equation}
In the code provided in this manuscript, this $\Tilde{R}^{\text{lat,W}}_{\rho\nu}(x)$ is used as an optional improved replacement of the finite-difference Ricci tensor $R^{\text{lat}}_{\rho\nu}(x)$ in the lattice action and in geometric observables.

\bibliographystyle{unsrt}
\bibliography{bibliography}

@article{tHooft:1974toh,
    author = "'t Hooft, Gerard and Veltman, M. J. G.",
    title = "{One-loop divergencies in the theory of gravitation}",
    doi = "10.1142/9789814539395_0001",
    journal = "Ann. Inst. H. Poincare Phys. Theor. A",
    volume = "20",
    number = "1",
    pages = "69--94",
    year = "1974"
}

@article{Goroff:1985sz,
    author = "Goroff, Marc H. and Sagnotti, Augusto",
    title = "{The Ultraviolet Behavior of Einstein Gravity}",
    reportNumber = "CALT-68-1289, LBL-19995, UCB-PTH-85-34",
    doi = "10.1016/0550-3213(86)90193-8",
    journal = "Nucl. Phys. B",
    volume = "266",
    pages = "709--736",
    year = "1986"
}

@article{Eichhorn:2022gku,
    author = "Eichhorn, Astrid and Schiffer, Marc",
    title = "{Asymptotic safety of gravity with matter}",
    eprint = "2212.07456",
    archivePrefix = "arXiv",
    primaryClass = "hep-th",
    month = "12",
    year = "2022"
}

@article{Reuter:1996cp,
    author = "Reuter, M.",
    title = "{Nonperturbative evolution equation for quantum gravity}",
    eprint = "hep-th/9605030",
    archivePrefix = "arXiv",
    reportNumber = "DESY-96-080",
    doi = "10.1103/PhysRevD.57.971",
    journal = "Phys. Rev. D",
    volume = "57",
    pages = "971--985",
    year = "1998"
}

@article{Percacci:2007,
    author = "Percacci, Roberto",
    title = "{Asymptotic Safety}",
    eprint = "0709.3851",
    archivePrefix = "arXiv",
    primaryClass = "hep-th",
    pages = "111--128",
    month = "9",
    year = "2007"
}

@article{Reuter:2012,
    author = "Reuter, Martin and Saueressig, Frank",
    title = "{Quantum Einstein Gravity}",
    eprint = "1202.2274",
    archivePrefix = "arXiv",
    primaryClass = "hep-th",
    doi = "10.1088/1367-2630/14/5/055022",
    journal = "New J. Phys.",
    volume = "14",
    pages = "055022",
    year = "2012"
}

@article{Eichhorn:2018yfc,
    author = "Eichhorn, Astrid",
    title = "{An asymptotically safe guide to quantum gravity and matter}",
    eprint = "1810.07615",
    archivePrefix = "arXiv",
    primaryClass = "hep-th",
    doi = "10.3389/fspas.2018.00047",
    journal = "Front. Astron. Space Sci.",
    volume = "5",
    pages = "47",
    year = "2019"
}

@article{Falls:2018ylp,
    author = {Falls, Kevin G. and Litim, Daniel F. and Schr{\"o}der, Jan},
    title = "{Aspects of asymptotic safety for quantum gravity}",
    eprint = "1810.08550",
    archivePrefix = "arXiv",
    primaryClass = "gr-qc",
    doi = "10.1103/PhysRevD.99.126015",
    journal = "Phys. Rev. D",
    volume = "99",
    number = "12",
    pages = "126015",
    year = "2019"
}

@article{Bonanno:2020bil,
    author = "Bonanno, Alfio and Eichhorn, Astrid and Gies, Holger and Pawlowski, Jan M. and Percacci, Roberto and Reuter, Martin and Saueressig, Frank and Vacca, Gian Paolo",
    title = "{Critical reflections on asymptotically safe gravity}",
    eprint = "2004.06810",
    archivePrefix = "arXiv",
    primaryClass = "gr-qc",
    doi = "10.3389/fphy.2020.00269",
    journal = "Front. in Phys.",
    volume = "8",
    pages = "269",
    year = "2020"
}

@article{Ambjorn:2000dv,
    author = "Ambjorn, Jan and Jurkiewicz, J. and Loll, R.",
    title = "{A Nonperturbative Lorentzian path integral for gravity}",
    eprint = "hep-th/0002050",
    archivePrefix = "arXiv",
    reportNumber = "AEI-2000-001, NBI-HE-00-03",
    doi = "10.1103/PhysRevLett.85.924",
    journal = "Phys. Rev. Lett.",
    volume = "85",
    pages = "924--927",
    year = "2000"
}

@article{Ambjorn:2001cv,
    author = "Ambjorn, Jan and Jurkiewicz, J. and Loll, R.",
    title = "{Dynamically triangulating Lorentzian quantum gravity}",
    eprint = "hep-th/0105267",
    archivePrefix = "arXiv",
    reportNumber = "AEI-2001-049",
    doi = "10.1016/S0550-3213(01)00297-8",
    journal = "Nucl. Phys. B",
    volume = "610",
    pages = "347--382",
    year = "2001"
}

@article{Ambjorn:2012jv,
    author = "Ambjorn, J. and Goerlich, A. and Jurkiewicz, J. and Loll, R.",
    title = "{Nonperturbative Quantum Gravity}",
    eprint = "1203.3591",
    archivePrefix = "arXiv",
    primaryClass = "hep-th",
    doi = "10.1016/j.physrep.2012.03.007",
    journal = "Phys. Rept.",
    volume = "519",
    pages = "127--210",
    year = "2012"
}

@article{Loll:2019rdj,
    author = "Loll, R.",
    title = "{Quantum Gravity from Causal Dynamical Triangulations: A Review}",
    eprint = "1905.08669",
    archivePrefix = "arXiv",
    primaryClass = "hep-th",
    doi = "10.1088/1361-6382/ab57c7",
    journal = "Class. Quant. Grav.",
    volume = "37",
    number = "1",
    pages = "013002",
    year = "2020"
}

@inproceedings{Ambjorn:2024CDT,
    author = "Ambj{\o}rn, J. and Loll, R.",
    title = "{Causal Dynamical Triangulations: Gateway to Nonperturbative Quantum Gravity}",
    eprint = "2401.09399",
    archivePrefix = "arXiv",
    primaryClass = "hep-th",
    month = "1",
    year = "2024"
}

@book{Eddington:1924,
  author    = {A. S. Eddington},
  title     = {The Mathematical Theory of Relativity},
  publisher = {Cambridge Univ. Press},
  year      = {1924}
}

@article{Einstein:1923,
title={The Theory of the Affine Field},
author={Albert B. Einstein},
journal={Nature},
year={1923},
volume={112},
pages={448-449},
url={https://doi.org/10.1038/112448a0}
}

@book{Schrodinger:1950, place={Cambridge}, series={Cambridge Science Classics}, title={Space-Time Structure}, publisher={Cambridge University Press}, author={Schrödinger, Erwin}, year={1985}, collection={Cambridge Science Classics}}

@article{Kijowski:2007,
    author = "Kijowski, Jerzy and Werpachowski, Roman",
    title = "{Universality of affine formulation in general relativity theory}",
    eprint = "gr-qc/0406088",
    archivePrefix = "arXiv",
    doi = "10.1016/S0034-4877(07)80001-2",
    journal = "Rept. Math. Phys.",
    volume = "59",
    pages = "1",
    year = "2007"
}

@Article{astronomy3010004,
AUTHOR = {Kijowski, Jerzy},
TITLE = {Gravity on a Large Scale—Does It Necessarily Look like It Does on a Small Scale?},
JOURNAL = {Astronomy},
VOLUME = {3},
YEAR = {2024},
NUMBER = {1},
PAGES = {29--42},
URL = {https://www.mdpi.com/2674-0346/3/1/4},
ISSN = {2674-0346},
DOI = {10.3390/astronomy3010004}
}

@article{Kalmykov:1994fm,
    author = "Kalmykov, M. Yu. and Pronin, P. I. and Stepanyantz, K. V.",
    title = "{Projective invariance and one loop effective action in affine metric gravity interacting with scalar field}",
    eprint = "hep-th/9408032",
    archivePrefix = "arXiv",
    doi = "10.1088/0264-9381/11/11/007",
    journal = "Class. Quant. Grav.",
    volume = "11",
    pages = "2645--2652",
    year = "1994"
}

@article{Kalmykov:1994yj,
    author = "Kalmykov, M. Yu. and Pronin, P. I.",
    title = "{The One loop divergences and renormalizability of the minimal gauge theory of gravity}",
    eprint = "hep-th/9412177",
    archivePrefix = "arXiv",
    doi = "10.1007/BF02113069",
    journal = "Gen. Rel. Grav.",
    volume = "27",
    pages = "873--886",
    year = "1995"
}

@article{Martellini:1983tx,
    author = "Martellini, M.",
    title = "{Renormalizability of quantum gravity wit a cosmological constant}",
    doi = "10.1103/PhysRevLett.51.152",
    journal = "Phys. Rev. Lett.",
    volume = "51",
    pages = "152--155",
    year = "1983"
}

@article{Martellini:1984ec,
    author = "Martellini, M.",
    title = "{Quantum gravity in the Eddington purely affine picture}",
    doi = "10.1103/PhysRevD.29.2746",
    journal = "Phys. Rev. D",
    volume = "29",
    pages = "2746--2755",
    year = "1984"
}

@article{Kijowski:1978,
    author = "Kijowski, Jerzy",
    title = "{On a new variational principle in general relativity and the energy of the gravitational field}",
    doi = "10.1007/BF00759646",
    journal = "Gen. Rel. Grav.",
    volume = "9",
    number = "10",
    pages = "857--877",
    year = "1978"
}

@article{Ferraris:1981,
    author = "Ferraris, M. and Kijowski, J.",
    title = "{On the equivalence of the relativistic theories of gravitation}",
    doi = "10.1007/BF00756921",
    journal = "Gen. Rel. Grav.",
    volume = "14",
    pages = "165--180",
    year = "1982"
}

@article{Poplawski:2007,
  author  = {N. J. Pop{\l}awski},
  title   = {A unified, purely affine theory of gravitation and electromagnetism},
  journal = {Int. J. Mod. Phys. D},
  volume  = {18},
  pages   = {809--839},
  year    = {2009},
  eprint  = {gr-qc/0701176}
}

@article{Poplawski:2008,
    author = "Poplawski, Nikodem J.",
    title = "{F(R) gravity in purely affine formulation}",
    eprint = "0706.4474",
    archivePrefix = "arXiv",
    primaryClass = "gr-qc",
    doi = "10.1142/S0217751X08039773",
    journal = "Int. J. Mod. Phys. A",
    volume = "23",
    pages = "1891--1901",
    year = "2008"
}

@article{Martellini:1984gy,
    author = "Martellini, M.",
    title = "{Quantum gravity in the Eddington purely affine picture}",
    doi = "10.1103/PhysRevD.29.2746",
    journal = "Phys. Rev. D",
    volume = "29",
    pages = "2746--2755",
    year = "1984"
}

@book{KijowskiTulczyjew:1979,
  title={A symplectic framework for field theories},
  author={Kijowski, Jerzy and Tulczyjew, Wlodzimierz M},
  year={1979},
  publisher={Springer}
}

@article{CervantesCota:2016,
    author = "Cervantes-Cota, Jorge L. and Liebscher, D. -E.",
    title = "{On constructing purely affine theories with matter}",
    eprint = "1607.04250",
    archivePrefix = "arXiv",
    primaryClass = "gr-qc",
    doi = "10.1007/s10714-016-2103-9",
    journal = "Gen. Rel. Grav.",
    volume = "48",
    number = "8",
    pages = "108",
    year = "2016"
}

@article{Vollick:2004,
  title={Born-Infeld-Einstein theory with matter},
  author={Vollick, Dan N},
  journal={Physical Review D—Particles, Fields, Gravitation, and Cosmology},
  volume={72},
  number={8},
  pages={084026},
  year={2005},
  publisher={APS}
}

@article{Banados:2010ix,
    author = "Banados, Maximo and Ferreira, Pedro G.",
    title = "{Eddington's theory of gravity and its progeny}",
    eprint = "1006.1769",
    archivePrefix = "arXiv",
    primaryClass = "astro-ph.CO",
    doi = "10.1103/PhysRevLett.105.011101",
    journal = "Phys. Rev. Lett.",
    volume = "105",
    pages = "011101",
    year = "2010",
    note = "[Erratum: Phys.Rev.Lett. 113, 119901 (2014)]"
}

@article{Avelino:2012ge,
    author = "Avelino, P. P.",
    title = "{Eddington-inspired Born-Infeld gravity: astrophysical and cosmological constraints}",
    eprint = "1201.2544",
    archivePrefix = "arXiv",
    primaryClass = "astro-ph.CO",
    doi = "10.1103/PhysRevD.85.104053",
    journal = "Phys. Rev. D",
    volume = "85",
    pages = "104053",
    year = "2012"
}

@article{BeltranJimenez:2017,
    author = "Beltran Jimenez, Jose and Heisenberg, Lavinia and Olmo, Gonzalo J. and Rubiera-Garcia, Diego",
    title = "{Born{\textendash}Infeld inspired modifications of gravity}",
    eprint = "1704.03351",
    archivePrefix = "arXiv",
    primaryClass = "gr-qc",
    reportNumber = "IFIC-17-23",
    doi = "10.1016/j.physrep.2017.11.001",
    journal = "Phys. Rept.",
    volume = "727",
    pages = "1--129",
    year = "2018"
}

@article{Chen:2016,
    author = "Chen, Che-Yu and Bouhmadi-Lopez, Mariam and Chen, Pisin",
    title = "{Modified Eddington-inspired-Born-Infeld Gravity with a Trace Term}",
    eprint = "1507.00028",
    archivePrefix = "arXiv",
    primaryClass = "gr-qc",
    doi = "10.1140/epjc/s10052-016-3879-1",
    journal = "Eur. Phys. J. C",
    volume = "76",
    pages = "40",
    year = "2016"
}

@article{Ambjorn:2020rcn,
    author = {Ambjorn, J. and Gizbert-Studnicki, J. and G{\"o}rlich, A. and Jurkiewicz, J. and Loll, R.},
    title = "{Renormalization in quantum theories of geometry}",
    eprint = "2002.01693",
    archivePrefix = "arXiv",
    primaryClass = "hep-th",
    doi = "10.3389/fphy.2020.00247",
    journal = "Front. in Phys.",
    volume = "8",
    pages = "247",
    year = "2020"
}

@article{Ambjorn:2024as,
  author  = {J. Ambj{\o}rn and J. Gizbert-Studnicki and A. G{\"o}rlich and J. Jurkiewicz and R. Loll},
  title   = {Is lattice quantum gravity asymptotically safe? Making contact between causal dynamical triangulations and the functional renormalization group},
  journal = {Phys. Rev. D},
  volume  = {110},
  pages   = {126006},
  year    = {2024},
  doi     = {10.1103/PhysRevD.110.126006},
  eprint  = {2408.07808},
  archivePrefix = {arXiv}
}

@book{Hamber:2009,
    author = "Hamber, Herbert W.",
    title = "{Quantum gravitation: The Feynman path integral approach}",
    doi = "10.1007/978-3-540-85293-3",
    publisher = "Springer",
    address = "Berlin",
    year = "2009"
}

@misc{silva_lqagmc_2025,
  author       = {A. Silva},
  title        = {Lattice Quantum Affine Gravity Monte Carlo Code},
  howpublished = {\url{https://github.com/silvaagu/LatticeQuantumAffineGravityMonteCarlo}},
  year         = {2025},
  note         = {Version v0.1.0, released 2025-12-01 (MIT License)},
}

@article{Ambjorn:1992EDT,
    author = "Ambjorn, Jan and Jurkiewicz, Jerzy",
    title = "{Four-dimensional simplicial quantum gravity}",
    reportNumber = "NBI-HE-91-60",
    doi = "10.1016/0370-2693(92)90709-D",
    journal = "Phys. Lett. B",
    volume = "278",
    pages = "42--50",
    year = "1992"
}

@article{Agishtein:1992EDT,
  title={Simulations of four-dimensional simplicial quantum gravity as dynamical triangulation},
  author={Agishtein, ME and Migdal, Alexander A},
  journal={Modern Physics Letters A},
  volume={7},
  number={12},
  pages={1039--1061},
  year={1992},
  publisher={World Scientific}
}

@article{Wilson:1974sk,
    author = "Wilson, Kenneth G.",
    editor = "Taylor, J. C.",
    title = "{Confinement of Quarks}",
    reportNumber = "CLNS-262",
    doi = "10.1103/PhysRevD.10.2445",
    journal = "Phys. Rev. D",
    volume = "10",
    pages = "2445--2459",
    year = "1974"
}

@book{ponomarev2017gauge,
  title={Gauge approach and quantization methods in gravity theory},
  author={Ponomarev, Vladimir Nikolaevi{\v{c}} and Barvinskij, Andrej Olegovi{\v{c}} and Obukhov, Yuri N},
  volume={10},
  year={2017},
  publisher={Nauka Moscow}
}

@article{Olmo:2011uz,
    author = "Olmo, Gonzalo J.",
    title = "{Palatini Approach to Modified Gravity: f(R) Theories and Beyond}",
    eprint = "1101.3864",
    archivePrefix = "arXiv",
    primaryClass = "gr-qc",
    doi = "10.1142/S0218271811018925",
    journal = "Int. J. Mod. Phys. D",
    volume = "20",
    pages = "413--462",
    year = "2011"
}

@article{Hehl:1994ue,
    author = "Hehl, Friedrich W. and McCrea, J. Dermott and Mielke, Eckehard W. and Ne'eman, Yuval",
    title = "{Metric affine gauge theory of gravity: Field equations, Noether identities, world spinors, and breaking of dilation invariance}",
    eprint = "gr-qc/9402012",
    archivePrefix = "arXiv",
    reportNumber = "TAUP-N192-94, TAUP-192-94",
    doi = "10.1016/0370-1573(94)00111-F",
    journal = "Phys. Rept.",
    volume = "258",
    pages = "1--171",
    year = "1995"
}

@article{Sotiriou:2006qn,
    author = "Sotiriou, Thomas P. and Liberati, Stefano",
    title = "{Metric-affine f(R) theories of gravity}",
    eprint = "gr-qc/0604006",
    archivePrefix = "arXiv",
    doi = "10.1016/j.aop.2006.06.002",
    journal = "Annals Phys.",
    volume = "322",
    pages = "935--966",
    year = "2007"
}

@article{Garcia-Parrado:2020lpt,
    author = "Garc{\'\i}a-Parrado, A. and Minguzzi, E.",
    title = "{Projective and amplified symmetries in metric-affine theories}",
    eprint = "2006.04040",
    archivePrefix = "arXiv",
    primaryClass = "gr-qc",
    doi = "10.1088/1361-6382/abed61",
    journal = "Class. Quant. Grav.",
    volume = "38",
    number = "13",
    pages = "135001",
    year = "2021"
}

@article{Ambjorn:2024bud,
    author = "Ambjorn, Jan and Gizbert-Studnicki, Jakub and Goerlich, Andrzej and Nemeth, Daniel",
    title = "{IR and UV limits of CDT and their relations to FRG}",
    eprint = "2411.02330",
    archivePrefix = "arXiv",
    primaryClass = "hep-lat",
    doi = "10.5506/APhysPolB.55.12-A2",
    journal = "Acta Phys. Polon. B",
    volume = "55",
    pages = "12--A2",
    year = "2024"
}

@article{Ambjorn:2022dvx,
    author = {Ambjorn, J. and Gizbert-Studnicki, J. and G{\"o}rlich, A. and N{\'e}meth, D.},
    title = "{Topology induced first-order phase transitions in lattice quantum gravity}",
    eprint = "2202.07392",
    archivePrefix = "arXiv",
    primaryClass = "hep-lat",
    doi = "10.1007/JHEP04(2022)103",
    journal = "JHEP",
    volume = "04",
    pages = "103",
    year = "2022"
}

@article{Ambjorn:2007jv,
    author = "Ambjorn, J. and Gorlich, A. and Jurkiewicz, J. and Loll, R.",
    title = "{Planckian Birth of the Quantum de Sitter Universe}",
    eprint = "0712.2485",
    archivePrefix = "arXiv",
    primaryClass = "hep-th",
    reportNumber = "ITP-UU-07-64",
    doi = "10.1103/PhysRevLett.100.091304",
    journal = "Phys. Rev. Lett.",
    volume = "100",
    pages = "091304",
    year = "2008"
}

@article{Ambjorn:2005db,
    author = "Ambjorn, J. and Jurkiewicz, J. and Loll, R.",
    title = "{Spectral dimension of the universe}",
    eprint = "hep-th/0505113",
    archivePrefix = "arXiv",
    reportNumber = "SPIN-05-05, ITP-UU-05-07",
    doi = "10.1103/PhysRevLett.95.171301",
    journal = "Phys. Rev. Lett.",
    volume = "95",
    pages = "171301",
    year = "2005"
}

@article{Ambjorn:2005qt,
    author = "Ambjorn, J. and Jurkiewicz, J. and Loll, R.",
    title = "{Reconstructing the universe}",
    eprint = "hep-th/0505154",
    archivePrefix = "arXiv",
    reportNumber = "SPIN-05-14, ITP-UU-05-18",
    doi = "10.1103/PhysRevD.72.064014",
    journal = "Phys. Rev. D",
    volume = "72",
    pages = "064014",
    year = "2005"
}

@article{Ambjorn:2014gsa,
    author = {Ambjorn, J. and G{\"o}rlich, A. and Jurkiewicz, J. and Kreienbuehl, A. and Loll, R.},
    title = "{Renormalization Group Flow in CDT}",
    eprint = "1405.4585",
    archivePrefix = "arXiv",
    primaryClass = "hep-th",
    doi = "10.1088/0264-9381/31/16/165003",
    journal = "Class. Quant. Grav.",
    volume = "31",
    pages = "165003",
    year = "2014"
}

@article{Knorr:2020bjm,
    author = "Knorr, Benjamin and Ripken, Chris",
    title = "{Scattering amplitudes in affine gravity}",
    eprint = "2012.05144",
    archivePrefix = "arXiv",
    primaryClass = "hep-th",
    doi = "10.1103/PhysRevD.103.105019",
    journal = "Phys. Rev. D",
    volume = "103",
    number = "10",
    pages = "105019",
    year = "2021"
}

@article{Percacci:2020bzf,
    author = "Percacci, Roberto",
    title = "{Towards Metric-Affine Quantum Gravity}",
    eprint = "2003.09486",
    archivePrefix = "arXiv",
    primaryClass = "gr-qc",
    doi = "10.1142/S0219887820400034",
    journal = "Int. J. Geom. Meth. Mod. Phys.",
    volume = "17",
    number = "supp01",
    pages = "2040003",
    year = "2020"
}

@article{Pagani:2015ema,
    author = "Pagani, Carlo and Percacci, Roberto",
    title = "{Quantum gravity with torsion and non-metricity}",
    eprint = "1506.02882",
    archivePrefix = "arXiv",
    primaryClass = "gr-qc",
    doi = "10.1088/0264-9381/32/19/195019",
    journal = "Class. Quant. Grav.",
    volume = "32",
    number = "19",
    pages = "195019",
    year = "2015"
}

@article{Gies:2022ikv,
    author = "Gies, Holger and Salek, Abdol Sabor",
    title = "{Asymptotically safe Hilbert{\textendash}Palatini gravity in an on-shell reduction scheme}",
    eprint = "2209.10435",
    archivePrefix = "arXiv",
    primaryClass = "hep-th",
    doi = "10.1140/epjc/s10052-023-11324-1",
    journal = "Eur. Phys. J. C",
    volume = "83",
    number = "2",
    pages = "173",
    year = "2023"
}

@article{Dona:2013qba,
    author = "Don{\`a}, Pietro and Eichhorn, Astrid and Percacci, Roberto",
    title = "{Matter matters in asymptotically safe quantum gravity}",
    eprint = "1311.2898",
    archivePrefix = "arXiv",
    primaryClass = "hep-th",
    doi = "10.1103/PhysRevD.89.084035",
    journal = "Phys. Rev. D",
    volume = "89",
    number = "8",
    pages = "084035",
    year = "2014"
}

@article{Bonanno:2015fga,
    author = "Bonanno, Alfio and Platania, Alessia",
    title = "{Asymptotically safe inflation from quadratic gravity}",
    eprint = "1507.03375",
    archivePrefix = "arXiv",
    primaryClass = "gr-qc",
    doi = "10.1016/j.physletb.2015.10.005",
    journal = "Phys. Lett. B",
    volume = "750",
    pages = "638--642",
    year = "2015"
}

@article{DAngelo:2023tis,
    author = "D'Angelo, Edoardo and Rejzner, Kasia",
    title = "{A Lorentzian Renormalization Group Equation for Gauge Theories}",
    eprint = "2303.01479",
    archivePrefix = "arXiv",
    primaryClass = "math-ph",
    doi = "10.1007/s00023-024-01535-x",
    journal = "Annales Henri Poincare",
    volume = "26",
    number = "12",
    pages = "4411--4459",
    year = "2025"
}

@article{DAngelo:2025yoy,
    author = {D'Angelo, Edoardo and Ferrero, Renata and Fr{\"o}b, Markus B.},
    title = "{De Sitter quantum gravity within the covariant Lorentzian approach to asymptotic safety}",
    eprint = "2502.05135",
    archivePrefix = "arXiv",
    primaryClass = "hep-th",
    doi = "10.1088/1361-6382/ade193",
    journal = "Class. Quant. Grav.",
    volume = "42",
    number = "12",
    pages = "125008",
    year = "2025"
}

@article{Saueressig:2025ypi,
    author = "Saueressig, Frank and Wang, Jian",
    title = "{Foliated asymptotically safe gravity: Lorentzian signature fluctuations from the Wick rotation}",
    eprint = "2501.03752",
    archivePrefix = "arXiv",
    primaryClass = "hep-th",
    doi = "10.1103/PhysRevD.111.106007",
    journal = "Phys. Rev. D",
    volume = "111",
    number = "10",
    pages = "106007",
    year = "2025"
}

@article{Fehre:2021eob,
    author = "Fehre, Jannik and Litim, Daniel F. and Pawlowski, Jan M. and Reichert, Manuel",
    title = "{Lorentzian Quantum Gravity and the Graviton Spectral Function}",
    eprint = "2111.13232",
    archivePrefix = "arXiv",
    primaryClass = "hep-th",
    doi = "10.1103/PhysRevLett.130.081501",
    journal = "Phys. Rev. Lett.",
    volume = "130",
    number = "8",
    pages = "081501",
    year = "2023"
}

@incollection{Weinberg:1979,
  author    = {S. Weinberg},
  title     = {Ultraviolet divergences in quantum theories of gravitation},
  booktitle = {General Relativity: An {E}instein Centenary Survey},
  editor    = {S. W. Hawking and W. Israel},
  publisher = {Cambridge Univ. Press},
  year      = {1979},
  pages     = {790--831}
}

@incollection{Morris:2022btf,
  author        = "Morris, Tim R. and Stulga, Dalius",
  title         = "{The Functional f(R) Approximation}",
  booktitle     = "{Handbook of Quantum Gravity}",
  editor        = "Bambi, Cosimo and Modesto, Leonardo and Shapiro, Ilya",
  publisher     = "Springer",
  address       = "Singapore",
  pages         = "1--33",
  year          = "2023",
  doi           = "10.1007/978-981-19-3079-9_19-1",
  eprint        = "2210.11356",
  archivePrefix = "arXiv",
  primaryClass  = "hep-th"
}

@incollection{Knorr:2022dsx,
  author        = "Knorr, Benjamin and Ripken, Chris and Saueressig, Frank",
  title         = "{Form Factors in Asymptotically Safe Quantum Gravity}",
  booktitle     = "{Handbook of Quantum Gravity}",
  editor        = "Bambi, Cosimo and Modesto, Leonardo and Shapiro, Ilya",
  publisher     = "Springer",
  address       = "Singapore",
  pages         = "1--49",
  year          = "2024",
  doi           = "10.1007/978-981-19-3079-9_21-1",
  eprint        = "2210.16072",
  archivePrefix = "arXiv",
  primaryClass  = "hep-th",
  reportNumber  = "NORDITA 2022-075"
}

@incollection{Martini:2022sll,
  author        = "Martini, Riccardo and Vacca, Gian Paolo and Zanusso, Omar",
  title         = "{Perturbative Approaches to Nonperturbative Quantum Gravity}",
  booktitle     = "{Handbook of Quantum Gravity}",
  editor        = "Bambi, Cosimo and Modesto, Leonardo and Shapiro, Ilya",
  publisher     = "Springer",
  address       = "Singapore",
  pages         = "1--46",
  year          = "2024",
  doi           = "10.1007/978-981-19-3079-9_25-1",
  eprint        = "2210.13910",
  archivePrefix = "arXiv",
  primaryClass  = "hep-th"
}

@incollection{Pawlowski:2023gym,
  author        = "Pawlowski, Jan M. and Reichert, Manuel",
  title         = "{Quantum Gravity from Dynamical Metric Fluctuations}",
  booktitle     = "{Handbook of Quantum Gravity}",
  editor        = "Bambi, Cosimo and Modesto, Leonardo and Shapiro, Ilya",
  publisher     = "Springer",
  address       = "Singapore",
  pages         = "1--70",
  year          = "2024",
  doi           = "10.1007/978-981-19-3079-9_17-1",
  eprint        = "2309.10785",
  archivePrefix = "arXiv",
  primaryClass  = "hep-th"
}

@incollection{Platania:2023srt,
  author        = "Platania, Alessia",
  title         = "{Black Holes in Asymptotically Safe Gravity}",
  booktitle     = "{Handbook of Quantum Gravity}",
  editor        = "Bambi, Cosimo and Modesto, Leonardo and Shapiro, Ilya",
  publisher     = "Springer",
  address       = "Singapore",
  pages         = "1--65",
  year          = "2023",
  doi           = "10.1007/978-981-19-3079-9_24-1",
  eprint        = "2302.04272",
  archivePrefix = "arXiv",
  primaryClass  = "gr-qc",
  reportNumber  = "NORDITA 2022-085"
}

@article{Reichert:2020mja,
    author = "Reichert, Manuel",
    title = "{Lecture notes: Functional Renormalisation Group and Asymptotically Safe Quantum Gravity}",
    doi = "10.22323/1.384.0005",
    journal = "PoS",
    volume = "384",
    pages = "005",
    year = "2020"
}

@article{Basile:2024oms,
    author = "Basile, Ivano and Buoninfante, Luca and Di Filippo, Francesco and Knorr, Benjamin and Platania, Alessia and Tokareva, Anna",
    title = "{Lectures in quantum gravity}",
    eprint = "2412.08690",
    archivePrefix = "arXiv",
    primaryClass = "hep-th",
    doi = "10.21468/SciPostPhysLectNotes.98",
    journal = "SciPost Phys. Lect. Notes",
    volume = "98",
    pages = "1",
    year = "2025"
}

@article{DAngelo:2023wje,
    author = "D'Angelo, Edoardo",
    title = "{Asymptotic safety in Lorentzian quantum gravity}",
    eprint = "2310.20603",
    archivePrefix = "arXiv",
    primaryClass = "hep-th",
    doi = "10.1103/PhysRevD.109.066012",
    journal = "Phys. Rev. D",
    volume = "109",
    number = "6",
    pages = "066012",
    year = "2024"
}

@article{Niedermaier:2006wt,
    author = "Niedermaier, Max and Reuter, Martin",
    title = "{The Asymptotic Safety Scenario in Quantum Gravity}",
    doi = "10.12942/lrr-2006-5",
    journal = "Living Rev. Rel.",
    volume = "9",
    pages = "5--173",
    year = "2006"
}

@article{Reuter:2001ag,
    author = "Reuter, M. and Saueressig, Frank",
    title = "{Renormalization group flow of quantum gravity in the Einstein-Hilbert truncation}",
    eprint = "hep-th/0110054",
    archivePrefix = "arXiv",
    reportNumber = "MZ-TH-01-27",
    doi = "10.1103/PhysRevD.65.065016",
    journal = "Phys. Rev. D",
    volume = "65",
    pages = "065016",
    year = "2002"
}

@article{Coumbe:2014nea,
    author = "Coumbe, Daniel and Laiho, John",
    title = "{Exploring Euclidean Dynamical Triangulations with a Non-trivial Measure Term}",
    eprint = "1401.3299",
    archivePrefix = "arXiv",
    primaryClass = "hep-th",
    doi = "10.1007/JHEP04(2015)028",
    journal = "JHEP",
    volume = "04",
    pages = "028",
    year = "2015"
}

@article{Ambjorn:2013eha,
    author = "Ambjorn, J. and Glaser, L. and Goerlich, A. and Jurkiewicz, J.",
    title = "{Euclidian 4d quantum gravity with a non-trivial measure term}",
    eprint = "1307.2270",
    archivePrefix = "arXiv",
    primaryClass = "hep-lat",
    doi = "10.1007/JHEP10(2013)100",
    journal = "JHEP",
    volume = "10",
    pages = "100",
    year = "2013"
}

@inproceedings{Schiffer:2025cqc,
    author = "Schiffer, Marc",
    title = "{Asymptotically safe quantum gravity: functional and lattice perspectives}",
    booktitle = "{24th International Conference on General Relativity and Gravitation (GR24) and 16th Edoardo Amaldi Conference on Gravitational (Amaldi16) Waves}",
    eprint = "2509.26352",
    archivePrefix = "arXiv",
    primaryClass = "gr-qc",
    month = "9",
    year = "2025"
}

@article{Ferraris:1982wci,
    author = "Ferraris, M. and Francaviglia, M. and Reina, C.",
    title = "{Variational formulation of general relativity from 1915 to 1925 {\textquotedblleft}Palatini's method{\textquotedblright} discovered by Einstein in 1925}",
    doi = "10.1007/BF00756060",
    journal = "Gen. Rel. Grav.",
    volume = "14",
    number = "3",
    pages = "243--254",
    year = "1982"
}


\end{document}